\documentclass[nonacm,sigplan]{acmart}

\usepackage{soul}

\usepackage{tikz}
\usepackage{amsmath}

\usepackage{filecontents}

\usepackage{graphicx}
\usepackage{textcomp}
\usepackage{xcolor}
\usepackage{float}
\usepackage{algorithm}
\usepackage{algpseudocode}

\usepackage{todonotes}
\usepackage[utf8]{inputenc}

\usepackage[normalem]{ulem}

\usepackage{amsmath}
\usepackage{amsfonts}
\usepackage{tabularx}
\usepackage{lscape}
\usepackage{mathtools}
\usepackage{booktabs}
\usepackage{color}
\usepackage{xcolor}
\usepackage{caption}
\usepackage{subcaption}
\usepackage{algpseudocode}
\usepackage{algorithm}

\usepackage{enumitem}
\usepackage[toc,page]{appendix}
\usepackage{array}
\usepackage{tikz}
\newcommand*\circled[1]{\tikz[baseline=(char.base)]{
            \node[shape=circle,draw,inner sep=0.8pt, minimum size=2pt] (char) {#1};}}
\newcommand{\rpoint}[1]{\circled{{\fontfamily{pcr}\selectfont\footnotesize{#1}}}}

\newcolumntype{L}[1]{>{\raggedright\let\newline\\\arraybackslash\hspace{0pt}}m{#1}}
\newcolumntype{C}[1]{>{\centering\let\newline\\\arraybackslash\hspace{0pt}}m{#1}}
\newcolumntype{R}[1]{>{\raggedleft\let\newline\\\arraybackslash\hspace{0pt}}m{#1}}

\usepackage{etoolbox}
\makeatletter
\patchcmd{\@makecaption}
  {\scshape}
  {}
  {}
  {}
\makeatother
\usepackage{comment}
\usepackage{multirow}
\usepackage{xspace}

\makeatletter
\newcommand*{\etc}{%
    \@ifnextchar{.}%
        {etc}%
        {etc.\@\xspace}%
}
\makeatother

\newcommand{\etal}{\textit{et al}. }

\def\BibTeX{{\rm B\kern-.05em{\sc i\kern-.025em b}\kern-.08em
    T\kern-.1667em\lower.7ex\hbox{E}\kern-.125emX}}

\begin{document}

\title{Evasive Hardware Trojan through Adversarial Power Trace}
\author{Behnam Omidi}
\email{bomidi@gmu.edu}
\affiliation{
  \institution{Geotge Mason University}
  \state{Virginia}
  \country{USA}
}

\author{Khaled N. Khasawneh}
\email{kkhasawn@gmu.edu}
\affiliation{
  \institution{Geotge Mason University}
  \state{Virginia}
  \country{USA}
}

\author{Ihsen Alouani}
\email{i.alouani@qub.ac.uk}
\affiliation{
  \institution{Queen's University Belfast}
  \state{ Northern Ireland}
  \country{UK}
}

\begin{abstract}
The globalization of the Integrated Circuit (IC) supply chain, driven by time-to-market and cost considerations, has made ICs vulnerable to hardware Trojans (HTs). Against this threat, a promising approach is to use Machine Learning (ML)-based side-channel analysis, which has the advatage of being a non-intrusive method, along with efficiently detecting HTs under golden chip-free settings.

In this paper, we question the trustworthiness of ML-based HT detection via side-channel analysis. We introduce a HT obfuscation (HTO) approach to allow HTs to bypass this detection method. Rather than theoretically misleading the model by simulated adversarial traces, a key aspect of our approach is the design and implementation of adversarial noise as part of the circuitry, alongside the HT. We detail HTO methodologies for ASICs and FPGAs, and evaluate our approach using TrustHub benchmark. Interestingly, we found that HTO can be implemented with only a single transistor for ASIC designs to generate adversarial power traces that can fool the defense with $100\%$ efficiency.  We also efficiently implemented our approach on a Spartan 6 Xilinx FPGA using 2 different variants: \textbf{(i)} DSP slices-based, and \textbf{(ii)} ring-oscillator-based design. Additionally, we assess the efficiency of countermeasures like spectral domain analysis, and we show that an adaptive attacker can still design evasive HTOs by constraining the design with a spectral noise budget. In addition, while adversarial training (AT) offers higher protection against evasive HTs, AT models suffer from a considerable utility loss, potentially rendering them unsuitable for such security application.

We believe this research represents a significant step in understanding and exploiting ML vulnerabilities in a hardware security context, and we make all resources and designs openly available online: \url{https://dev.d18uu4lqwhbmka.amplifyapp.com}

\end{abstract}

\maketitle 
\pagestyle{plain} 

\section{Introduction}

\begin{figure*}[!ht]
  \includegraphics[width=\textwidth]{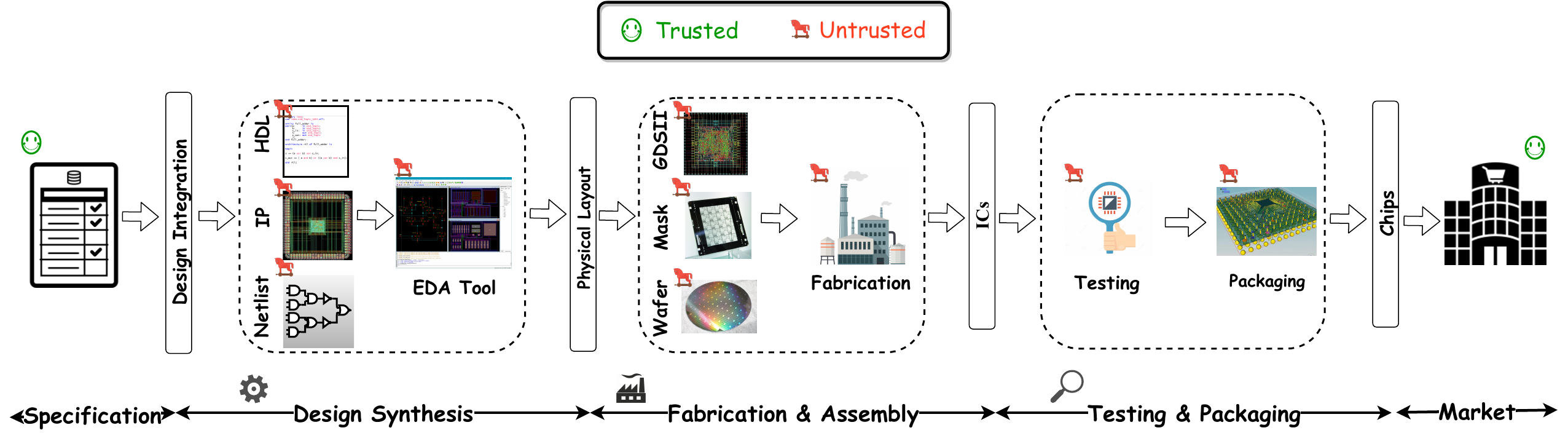}
  \caption{The credibility level of IC production life-cycle}
  \label{fig:ic_production_life_cycle}
\end{figure*}

The increasing demand for shorter time-to-market and cost-effective hardware design and manufacturing has globalized the IC supply chain, exposing ICs to hardware-based security threats like counterfeiting, IP piracy, reverse engineering, and HTs due to outsourcing to untrusted entities. The modern IC production life-cycle can be divided into three main phases: \textbf{\textit{design, fabrication, testing}} \cite{tehranipoor2010survey}. HTs are malicious modifications of a circuit to control, modify, disable, monitor, or affect the operation of the circuit. Figure \ref{fig:ic_production_life_cycle} illustrates the standard SoC design life cycle, from system-level specification to fabrication. Various stages in the IC supply chain are potential points for Hardware Trojan (HT) injection. This could be a malicious in-house designer or team can alter the hardware design at different abstraction levels, while relying on third party IPs or outsourcing manufacturing facilities could create a vulnerability for adversaries to tamper with and infect the design. 


Underestimating HT threats can lead to severe consequences. For example, F-Secure's 2020 report on counterfeit Cisco switches highlighted the risks of HTs in bypassing authentication processes~\cite{fsecure2020}. There have been other instances, like Bloomberg's claims in 2018 and 2021 about compromised servers in major companies' data centers, although these were disputed~\cite{bloomberg}. Additionally, the 2008 bombing of a Syrian nuclear facility after a backdoor disabled its radar potentially due to a HT~\cite{syrianRadar}. These threats pose significant risks to various critical systems, making the integrity of the semiconductor supply chain a pressing concern globally.

Researchers have explored various methods to detect HTs in chip design and manufacturing. These include destructive and non-destructive techniques~\cite{chakraborty2009hardware}. Destructive techniques involve dissecting chips to reverse engineer their layers~\cite{bao2015reverse}. Although these methods can be highly accurate, they are costly, time-consuming, and render the chip unusable, limiting their practicality to a single chip analysis. In contrast, non-destructive methods focus on the chip's functionality or side-channel signals, offering a  more practical alternatives~\cite{zhou2015cost}. These include Logic Testing~\cite{saha2015improved}, IP Trust Verification~\cite{xue2020ten}, Optical Detection~\cite{dong2020hardware}, Side-Channel Analysis~\cite{nguyen2020novel}, and ML-based Detection~\cite{yang2021side}. Particularly, the work in \cite{faezi2021htnet} presented a ML-based approach for HT detection using power side channel traces. This is particularly interesting post-silicon approach because it does not require golden-chip, beyond the fact that it is non-intrusive and highly efficient \cite{faezi2021htnet}.

 In this paper, we question the trustworthiness of ML-based HT detection using power side-channel data. 

 \noindent\textbf{Research challenges.} While the vulnerability of ML models to adversarial noise is established in the literature, generating \textbf{adversarial circuits}, i.e., adversarial noise implementable in circuitry, is a novel problem. In fact, the model's input in \cite{faezi2021htnet} is a power trace, and so is the adversarial noise generated by existing methods. Therefore, our first challenge is \textbf{(C1)}: given an adversarial power trace, \textit{how to generate an adversarial circuit that consumes this trace?} Moreover, given the threat model of HT detection, the second challenge \textbf{(C2)} is that \textit{adversarial circuits} need to be automatically designed with the lowest possible resource utilisation. Finally, the third challenge \textbf{(C3)} is how to design defenses and to which extent these defenses can be bypassed by adaptive attacks.
 
 \noindent\textbf{Attack.} We propose HTO, a method that exploits the vulnerability of ML to adversarial noise to generate evasive HTs that are undetectable by ML-based approaches. We first generate a targeted adversarial patch that takes hardware noise into account and successfully fools a state-of-the-art HT defense approach. We then propose new approaches for designing obfuscation circuits that consume adversarial power traces for both ASICs and FPGAs. To make the attack even stealthier, we optimize the generated adversarial noise to minimize the required circuitry to obfuscate the HT.

We evaluate HTO using the TrustHub benchmark with a recent ML-based HT detection technique and show that our approach efficiently bypasses the defense. We also demonstrate that HTO can be implemented with \textbf{a single transistor} for ASIC designs, making it very stealthy. 

\noindent\textbf{Countermeasures and adaptive attack.} For a comprehensive security evaluation, we investigate defenses and propose an adaptive attack. We propose a spectral defense in the \textit{frequency domain} to improve the robustness of ML-based detection and investigate an adaptive attack that generates noise in specific frequency range to bypass this defense. Additionally, we examine AT as a countermeasure against HTO, but find it can be circumvented with increased attacker power budget and results in reduced baseline accuracy.


To the best of our knowledge, this is the first work to practically exploit ML vulnerabilities to design adaptive hardware security attacks.

\noindent\textbf{Contributions.} The key contributions of this paper can be summarized as follows:

\noindent\textbf{Evasive HT.} We propose and develop an adversarial power generation methodology that allows HTs to bypass ML-based side channel defense. We design configurable adversarial patch generation circuits for various platforms (ASIC and FPGA) to consume \textbf{adversarial} power during the activation of HT to bypass the power-analysis detection. (Section \ref{sec:APTGen}, \ref{sec:APTImp})\\
\noindent\textbf{Attack optimisation.} We optimize the generated noise to achieve minimal resource utilisation and showed that an effective HTO design can be implemented with only $1$ Transistor in an ASIC. (Section \ref{sec:HTO_opt}) \\
\noindent\textbf{Countermeasures and adaptive attacks.} We propose a spectral domain countermeasure for HTO (Section \ref{sec:counter}) and explore an adaptive attack by considering the a spectral domain adversarial noise budget (Section \ref{sec:adaptive}). We also explore the practicality of adversarial training taking into account the utility constraints of the security use case.

\section{Threat Model}

In this paper, we consider that the adversary's objective is to \emph{evade post-silicon power-based ML detection of the HT}. 

\noindent\textbf{Attacker's capabilities:} The attacker within this threat model has the capability to modify the design and insert HTs at any point in the victim circuit, as commonly assumed in the state-of-the-art~\cite{clements2018hardware, huang2020survey, faezi2021brain,faezi2021htnet}. They can inject additional circuitry, but cannot alter the defender's ML-based HT detector, which is trained to detect the existence of a triggered HT through power analysis.\\
\noindent\textbf{Defender:} We assume the defender uses a ML-based HT detector which analyzes the circuit power trace. We specifically focus on HTnet~\cite{faezi2021htnet} as a golden chip-free technique, and which is trained by the power signals gathered on TrustHub benchmarks \cite{9fwb-8978-21} to detect HTs, due to its superior detection performance. \\
\noindent\textbf{Attacker's knowledge:} We assume that the attacker is aware of the defender's use of this ML model but sees it as a black-box; unable to modify its parameters or directly interfere with its operation.\\
\noindent\textbf{Objective:} An adversary, using information learnt about the ML-based defense mechanism, trains a proxy model and tries to craft power trace perturbations added to the ML input to force it to output wrong labels. 




\section{Adversarial Power Trace Generation}
\label{sec:APTGen}
In this section, we introduce the process of generating an adversarial patch which is a sneaky power trace augmented to the circuit power to force the ML-based HT detector model to wrongly label the HT inserted ICs. We precise it is an adversarial \textbf{patch} because the noise needs to be "\textit{universal}", i.e., the adversarial noise should fool the model regardless of the specific sample. This is in contrary to classical adversarial noise which is crafted for a given input sample. In the following, we use adversarial patch interchangeably with adversarial noise. 


\noindent\textbf{Problem formulation:}  Given an original input power trace $x$ and a victim classification model $f(.)$, the problem of generating an adversarial example $x_{adv}$ can be formulated as a constrained optimization:
\begin{equation}
\small
\begin{split}
        C(x + \delta) \neq C(x), ~ \forall ~ x \sim \mu_{HT}, and \\
        \left\Vert \delta \right\Vert = \mathcal{D}(x, x+\delta) \leq \varepsilon 
\end{split}
\end{equation}
Where $\mu_{HT}$ is the data distribution corresponding to triggered hardware-Trojans, which is our target for a classifier $C(.)$ to be fooled, i.e., evading detection. This means that we want to generate one noise ($\delta$)
$\varepsilon$ controls the magnitude of the adversarial patch.


\noindent
{\bf Distance Metrics:}
For evading the detection, the adversarial perturbations should be kept minimal. To measure the magnitude of adversarial noise,  $L_p$ metrics are generally used in the literature \cite{asplos, fgsm,CW}. In our case, the samples are power  traces and $L_\infty$ is more appropriate to measure the detectability of a potential power trace since $L_\infty$ measures the maximum difference for all elements at corresponding positions in the two samples. 

\noindent
{\bf Adversarial Patch Generation:}
We generate the adversarial power noise by solving the optimization problem and considering the detection loss function. Given a batch of power traces $X_i$ and corresponding target labels $\ell$. In each iteration, the algorithm adds the precalculated perturbation to the Trojan inserted trace and adjusts the new noise along the direction of its gradient w.r.t the inference loss $\partial J/\partial x$. Algorithm \ref{Algo:patchGen} describes the generation of adversarial power noise process.

\begin{algorithm}[!ht]
\small
\caption{Adversarial Power Trace Generation.}
\label{Algo:patchGen}
\begin{algorithmic}[1]
\State \textbf{Input:} Data set: $\mathcal{D}$, classifier: $C$,  noise magnitude: $\varepsilon$, standard deviation: $\sigma$, number of iterations: $N$.

\State \textbf{Output:} $\delta$ Adversarial power trace.
\State Initialize $\delta \leftarrow rand()$
\While{$iter < N$ }
    \For{each Batch $X_i \sim \mathcal{D}$}
        \For{all $x_{j}$ s.t. $C(x_{j})=1$} \Comment{ \textcolor{gray}{//HT distribution.}}\\
            \State $\delta \leftarrow \frac{1}{|X_i|} \sum_{j=1}^{|X_i|}
             \{ \{\alpha sign(\nabla_{x} J_{\theta}(C({x_{j} + \delta }),\ell))\}\}$ \\
            \State $\delta \leftarrow Clip_\varepsilon\{ \delta \} + z \sim N\left(0, \sigma^{2}\right)$
        \EndFor
    \EndFor
\State $iter += 1$
\EndWhile
\end{algorithmic}
\end{algorithm}

Where the attack strength is determined by $\alpha$ which is the perturbation constraint, $C({x_{j} + \delta})$ computes the output of ML based classifier, $J_{\theta}$ is the cross entropy loss function, and the sign function is indicated as $sgn(·)$. To produce a universal patch for traces in dataset, average of all generated noises has been forwarded to the next iteration. Since the negative values for the patch do not have any physical meaning in terms of power consumption and to ensure the the perturbation is restricted to the specified power budget $\epsilon$, The attack is followed by a clipping operation ($ 0 \leq \delta \leq \epsilon$).  A Gaussian noise $z$ with the mean value of $0$ and standard deviation $\sigma$ has been added to the perturbation to add robustness to the measurement and environment noise.

\begin{figure}[!ht]
\centering
	\resizebox{\linewidth}{!}{\includegraphics{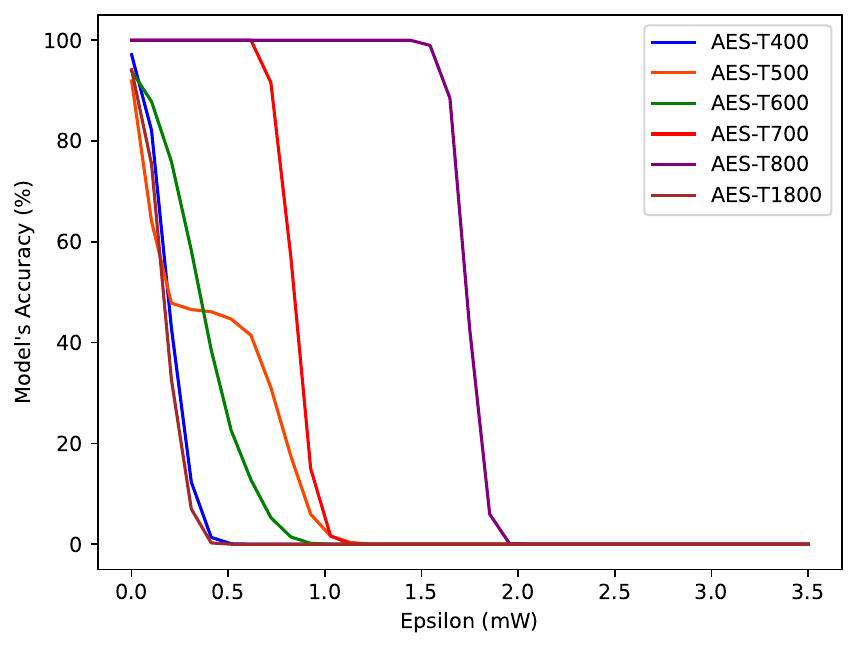}}
	\caption{HT detection accuracy based on changing the maximum power budget
     }
       \label{fig:model_accuracy_for_synch_unsynch} 
\end{figure}

Based on the described algorithm, the effectiveness of different power budgets on hardware-Trojan ML detector accuracy has been evaluated in Figure \ref{fig:model_accuracy_for_synch_unsynch}. The figure shows that for fooling the ML model with $100\%$ efficiency we can restrict the maximum power of adversarial patches to $3mW$.

\section{HTO: Proposed Adversarial Circuit Design}
\label{sec:APTImp}
In the previous section, we generate an adversarial patch that is able to fool a power side-channel based HT detection system. However, the adversarial noise we consider in Section \ref{sec:APTGen} is in the input space, i.e., it is a power trace. In a practical scenario, we cannot act on the power trace directly, as stated earlier in our threat model. 
In this section, we propose a methodology to design a circuitry that consumes a given power trace, which is the adversarial power trace generated in Section \ref{sec:APTGen}. Our objective is to answer the following question: \textbf{Given a target (adversarial) power trace, how to design a circuit that consumes a trace that emulates the target trace?}


In the following, we propose a design methodology for both FPGA and ASIC platforms. We implement a transistor-based circuit for ASIC and (i) RO-based as well as (ii) a DSP-based circuit design for FPGA to emulate the adversarial patch on both platforms.

\subsection{HTO Circuit Design for ASICs}

\begin{figure}[tp]
    \centering
	\includegraphics[width=\linewidth, height=6cm]{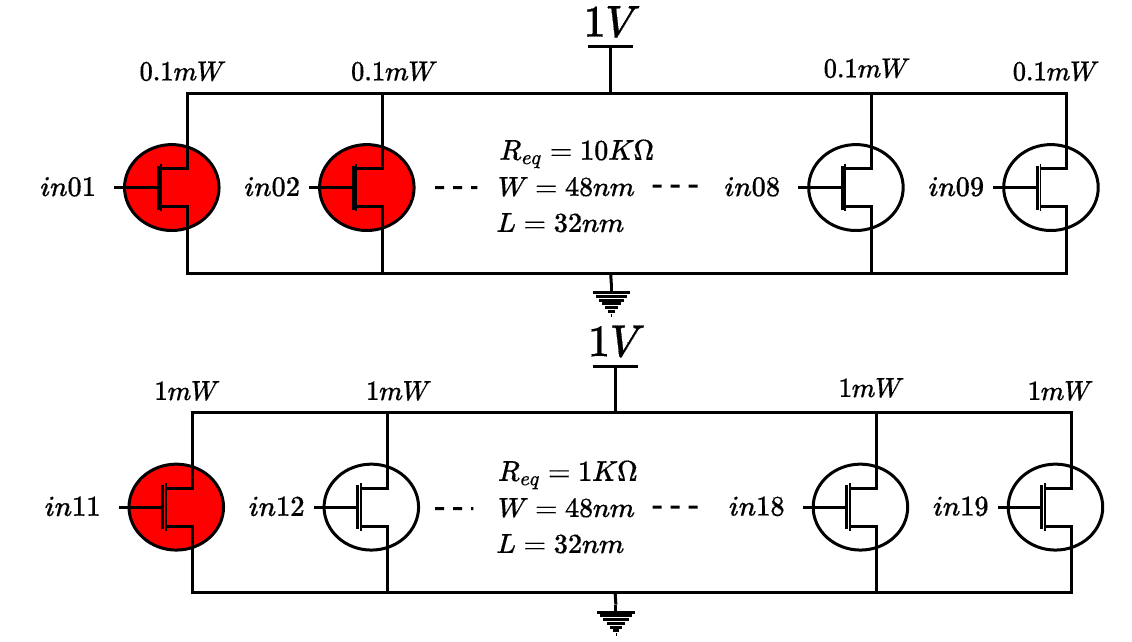}
	\caption{ASIC adversarial power network example}
	\label{fig:ASCI_nois_power_example} 
\end{figure}

For designing a circuit to generate adversarial power noise in ASIC, we propose a configurable power consumption unit. Specifically, the configurable power consumption unit is a network of intrinsic resistors hidden in transistors which are able to configure by activating transistors. The hidden resistors can be specified by the feature parameters of transistors during the design phase. Based on the samples in the adversarial power noise generated by the algorithm \ref{Algo:patchGen}, the equivalent hidden resistors are calculated, and corresponding vectors are activated in the power network. An example of this conversion for $P = 1.2mW$ and $V=1v$ with $18$ transistors has been shown in Figure \ref{fig:ASCI_nois_power_example} and algorithm \ref{algth:vector_generator}.


\begin{algorithm}[!ht]
\small
\caption{Configuration Vector Generator}
\label{algth:vector_generator}
\begin{algorithmic}[1]
\State \textbf{Input:} Noise samples: $\mathcal{N}$

\State \textbf{Output:} Configuration vectors: $\mathcal{V}$ 

\State Initialize $\mathcal{V} \leftarrow 0$

\For{each noise sample $P_i$ in $\mathcal{N}$}
    \For{each digit $d_l$ in $P_i$}
        \For{each cell $c_j$ in a raw of circuit}
            \If{$j < d_l$}
                \State   $\mathcal{V}_{lj} = 1$
            \Else
                \State   $\mathcal{V}_{lj} = 0$
            \EndIf
        \EndFor
    \EndFor
\EndFor

\end{algorithmic}
\end{algorithm}

\begin{figure}[tp]
    \centering
    \includegraphics[width=\linewidth]{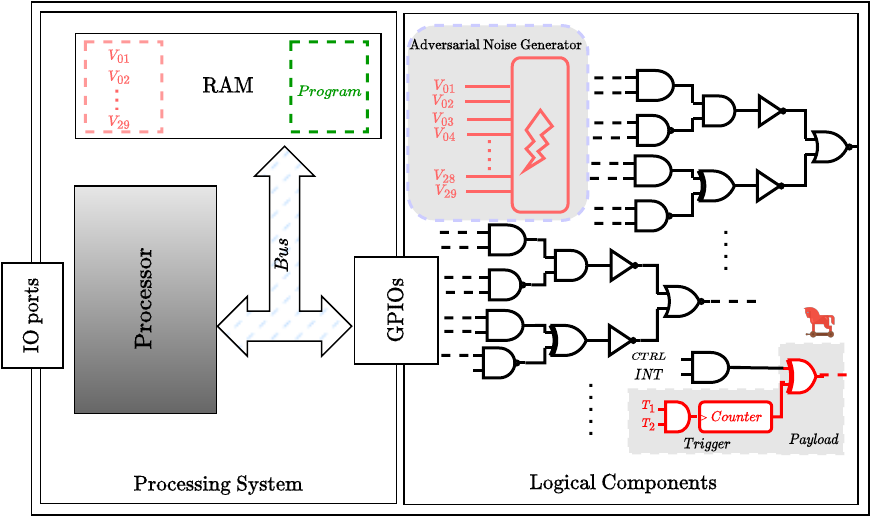}
    \caption{Illustration of a configuration vector generator in a SoC case}
    \label{fig:SoC_system_vector_generator} 
\end{figure}

There are multiple ways to store configuration vectors: 
\begin{itemize}
    \item Vectors can be stored in a Block-RAM where the order of vectors reflects the consumed power in each cycle by the HTO.
    \item For a System on Chip (SoC) circuit which has a processor, it is recommended to store the vectors in the processor's RAM to minimize resource utilization. The vectors can feed to the configurable power network by a port connected to the processor's bus. The structure of stored vectors in the SoC system has been shown in Figure \ref{fig:SoC_system_vector_generator}.
\end{itemize}

\subsection{HTO Circuit Design for FPGAs}

\begin{figure}[!ht]
    \centering
    \captionsetup[sub]{font=Large}
    \resizebox{\linewidth}{!}{
    \begin{subfigure}[b]{\linewidth}
    \centering
	\includegraphics{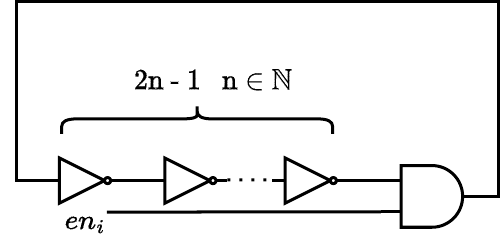}
	\caption{Conventional Ring Oscillator}
	\label{fig:conventional_ring_oscillator} 
    \end{subfigure}
    
    \begin{subfigure}[b]{\linewidth}
    \centering
	\includegraphics{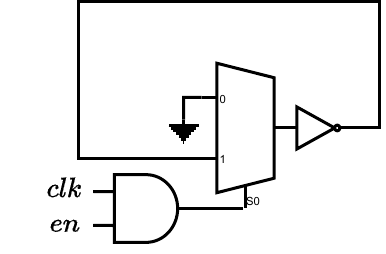}
	\caption{Synchronized Ring Oscillator}
	\label{fig:synch_ring_oscillator} 
    \end{subfigure}}
    \caption{Ring Oscillator}
    \label{fig:ring_oscillator}
\end{figure}

Not having access to the parameters of transistors, we take a different approach to design an HTO circuit in the FPGA platform. We propose two different logical circuits, which are ring oscillator-based and DSP-based networks to consume adversarial power. The implementation details of each circuit are discussed as follows: 

\noindent
\textbf{1- Ring Oscillators:} 
A network of configurable ring oscillators illustrated in Figure \ref{fig:ring_oscillator} has been proposed to consume generated adversarial power traces on FPGA. The ring oscillator frequency of the oscillation formula is:
\begin{equation}
\small
\label{eq:adv}
     \begin{array}{rlc}
   f= \frac{1}{2n\tau}
\end{array}
\end{equation}

Where $\tau$ is the propagation delay for a single inverter, and $n$ is the number of inverters which should be an odd number. For generating a sneaky adversarial power, the number of inverters should be as minimum as possible (one inverter) leading to the maximum frequency and power consumption. Unfortunately, our experimental setup can not capture the power consumed with the conventional ring oscillator illustrated in Figure \ref{fig:conventional_ring_oscillator}. Therefore, we modified our ring oscillator design to the one shown in Figure \ref{fig:synch_ring_oscillator} to synchronize our circuit with the whole system and capture the power precisely. In the new design, by logical-anding the enable signal with the FPGA clock, we synchronize the capture power devices with the clock generated by the ring oscillator. We can consume around $1 mW$ by using two ROs in our experimental setup limiting our resolution to $1 mW$.

\begin{figure}[!ht]
    \centering
	\includegraphics[width=0.8\linewidth]{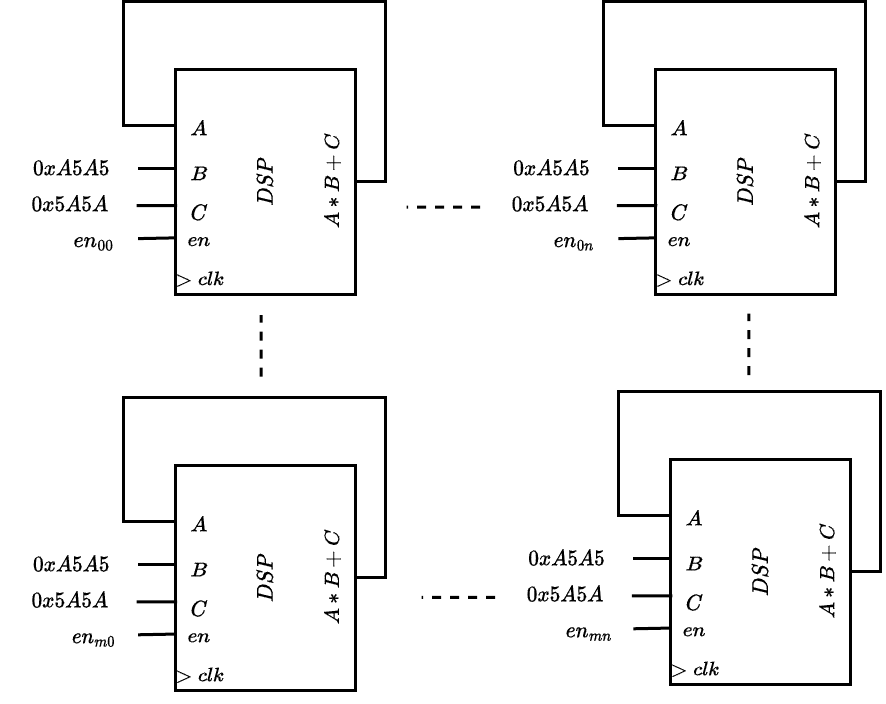}
	\caption{DSP-based network design}
	\label{fig:DSP_Network} 
\end{figure}

\noindent
\textbf{2- DSPs:} 
 We Exploit DSP components in FPGA to implement an obfuscated design that evades hardware-Trojan detection techniques. We notice that by a DSP with configuration in Figure \ref{fig:DSP_Network}, we can consume $1mW$ power in each clock cycle. Based on our results, we can achieve $100\%$ efficiency to fool ML classifier defense by using at least $3$ DSPs.
 
\noindent
\textbf{Configuration vectors:} 
Similar to the ASIC design, the configuration vectors can be stored in Block-RAMs or for an SoC circuit, vectors can be stored in the processor's RAM and sent to a general I/O port connected to the HTO circuit.

\subsection{Experimental setup}

\begin{figure}[!ht]
    \centering
	\includegraphics[width=\linewidth]{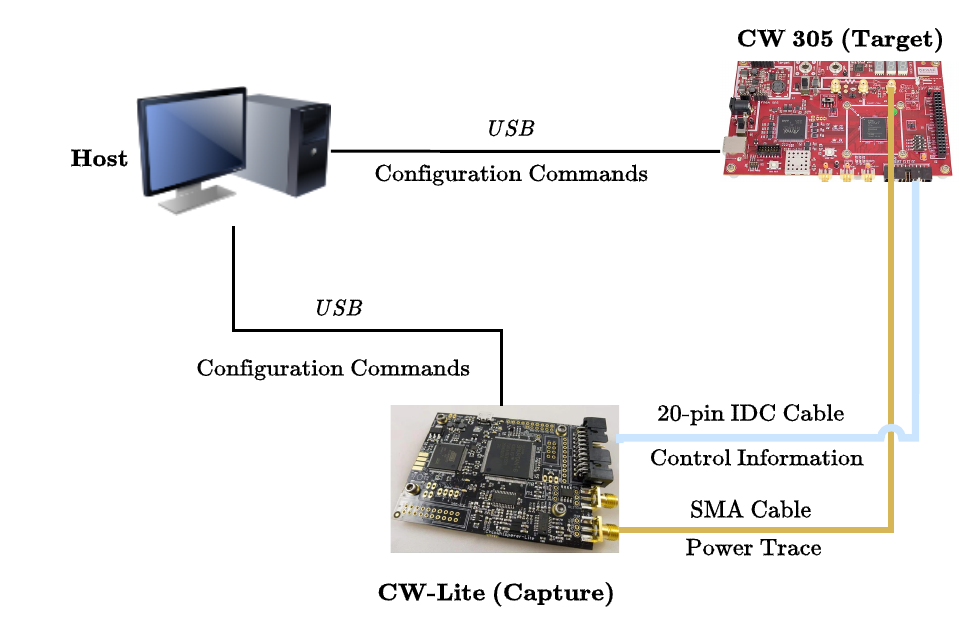}
	\caption{FPGA setup}
	\label{fig:fpga_power_trace_setup} 
\end{figure}

Different experimental setups have been used for the ASIC and FPGA platforms to generate the adversarial power traces. In ASIC, the adversarial patches generated to fool the classifier were simulated by $HSpice$ $2019$ with $32nm$ low power $PTM$. Generation of the equivalent circuit with appropriate configuration vectors in $HSpice$ format for each patch has been automated by a $Python$ code. For capturing the power in FPGA, we leverage ChipWhisperer 305 with Xilinx spartan 6 as a target device and ChipWhisperer-lite as a capture device. As shown in the Figure \ref{fig:fpga_power_trace_setup} the control of Artix-7 and Spartan-6 on FPGA manage by a USB connection from the laptop. Control information of our power measurements sends through the 20-pin connector connected to ChipWhisperer-lite. The power measurements will be done through an SMA cable to the ChipWhisperer-lite and sent to the laptop through a USB connection.

\subsection{Results}

Results on ASIC illustrate the negligible difference between the generated noises and simulated ones with the mean square error of $0.0002$. The power consumption resolution on FPGA restricts us to only implementing the patches with the granularity of $1mW$. Hence, The accuracy of the simulation drops in a way that the value of $0.31$ is reported for mean square error. Tables \ref{table:ASIC_synchronized} and \ref{table:FPGA_synchronized} illustrate the resource utilization and model’s accuracy on ASIC and FPGA respectively. To drop the model's accuracy to $0\%$, it requires the maximum of $4$ ring-oscillators to consume power in the range $0mW$ to $2mW$ with the resolution of $1mW$ on FPGA. However, the maximum of $12$ transistors need to generate the power in the range $0mW$ to $2mW$ with the resolution of $0.1mW$ to achieve $100\%$ efficiency in misclassification for ASIC. 

\begin{table}[!ht]
\small
    \centering
    \resizebox{\linewidth}{!}{
    \begin{tabular}{| C{1.7cm} | C{2cm} | C{1.5cm}| C{2cm} | C{1.5cm}|}
             \hline
             \multirow{3}{1.7cm}{\centering Dataset}&\multirow{3}{2cm}{\centering HT detection accuracy (\%)} &\multirow{3}{1.5cm}{\centering \# of transistors} &  \multirow{3}{2cm}{ \centering HT detection accuracy after patching (\%)} & \multirow{3}{1.5cm}{\centering Noise Budget (mW)}\\
             &&&&\\
             &&&& \\\hline\hline

             \multirow{1}{1.7cm}{\centering AES-T400} & \multirow{1}{2cm}{\centering 97.1} & 6  & 0 & \multirow{1}{1.5cm}{\centering 0-0.6} \\\hline

             \multirow{1}{1.7cm}{\centering AES-T500} & \multirow{1}{2cm}{\centering 91.91} & 11  & 0 & \multirow{1}{1.5cm}{\centering 0-1.2} \\\hline

             \multirow{1}{1.7cm}{\centering AES-T600} & \multirow{1}{2cm}{\centering 93.87} & 11 & 0 & \multirow{1}{1.5cm}{\centering 0-1} \\\hline

             \multirow{1}{1.7cm}{\centering AES-T700} & \multirow{1}{2cm}{\centering 100} & 11  & 0 & \multirow{1}{1.5cm}{\centering 0-1.2} \\\hline

             \multirow{1}{1.7cm}{\centering AES-T800} & \multirow{1}{2cm}{\centering 100} & 12 & 0 & \multirow{1}{1.5cm}{\centering 0-2} \\\hline

             \multirow{1}{1.7cm}{\centering AES-T1800} & \multirow{1}{2cm}{\centering 94.12} & 5  & 0 & \multirow{1}{1.5cm}{\centering 0-0.5} \\\hline               
    \end{tabular}
    }
    
    \caption{Resource utilization and HT detection accuracy on ASIC}
    \label{table:ASIC_synchronized}
\end{table}

\begin{table}[!ht]
\small
    \centering
    \resizebox{\linewidth}{!}{
    \begin{tabular}{| C{1.7cm} | C{2cm} | C{1cm} | C{1.5cm}| C{2cm}   | C{1.5cm}|}
             \hline
             \multirow{3}{1.7cm}{\centering Dataset} & \multirow{3}{2cm}{\centering HT detection accuracy (\%)} & \multirow{3}{1cm}{\centering Method} & \multirow{3}{1.5cm}{\centering \# of CLBs} & \multirow{3}{2cm}{ \centering HT detection accuracy after patching (\%)} & \multirow{3}{1.5cm}{\centering Noise Budget (mW)}\\
             &&&&&\\
             &&&&&\\\hline\hline

             \multirow{2}{1.7cm}{\centering AES-T400} & \multirow{2}{2cm}{\centering 97.1} & RO & 2 &  0 & \multirow{2}{1.5cm}{\centering 0-1} \\\cline{3-5}
             && DSP & 1  & 0 & \\\hline

             \multirow{2}{1.7cm}{\centering AES-T500} & \multirow{2}{2cm}{\centering 91.91} & RO & 4  & 0 & \multirow{2}{1.5cm}{\centering 0-2} \\\cline{3-5}
             && DSP & 2 & 0 & \\\hline

             \multirow{2}{1.7cm}{\centering AES-T600} & \multirow{2}{2cm}{\centering 93.87} & RO & 2  & 0 & \multirow{2}{1.5cm}{\centering 0-1} \\\cline{3-5}
             && DSP & 1  & 0 & \\\hline

             \multirow{2}{1.7cm}{\centering AES-T700} & \multirow{2}{2cm}{\centering 100} & RO & 4  & 0 & \multirow{2}{1.5cm}{\centering 0-2} \\\cline{3-5}
             && DSP & 2  & 0 & \\\hline

             \multirow{2}{1.7cm}{\centering AES-T800} & \multirow{2}{2cm}{\centering 100} & RO & 4  & 0 & \multirow{2}{1.5cm}{\centering 0-2} \\\cline{3-5}
             && DSP & 2  & 0 & \\\hline

             \multirow{2}{1.7cm}{\centering AES-T1800} & \multirow{2}{2cm}{\centering 94.12} & RO & 2  & 0 & \multirow{2}{1.5cm}{\centering 0-1} \\\cline{3-5}
             && DSP & 1  & 0 & \\\hline               
    \end{tabular}
    }
    
    \caption{Resource utilization and HT detection accuracy on FPGA}
    \label{table:FPGA_synchronized}
\end{table}

\textbf{Discussion--} Our results show that the generated HTO consumes power traces that accurately emulate the adversarial patch impact on the victim system. However, the high resource utilization required by the HTO is a clear limitation of its threat, which can be easily exposed by state-of-the-art techniques. In the following, we will investigate optimisation techniques to reduce the resource utilisation for more sneaky HTO designs.



\section{HTO Optimization}
\label{sec:HTO_opt}
In this section, we tackle the challenge of reducing resource utilization of HTO circuits for a more practical attack. We start by analyzing the adversarial patch values distribution. We observe that the noise power distribution is biased towards a relatively limited number of values. Using this observation, we adapted the adversarial power trace generation algorithm to have constraints on their values within a target space, thereby producing \textbf{quantized} power traces which require lower resource utilisation to emulate by HTOs. 

\subsection{Adversarial patch values distribution} \label{sec:patchdist}

\begin{figure}[!ht]
        \captionsetup[sub]{font=huge}
	\resizebox{\linewidth}{!}{
            \begin{subfigure}[b]{\linewidth}
                \includegraphics[width=\linewidth]{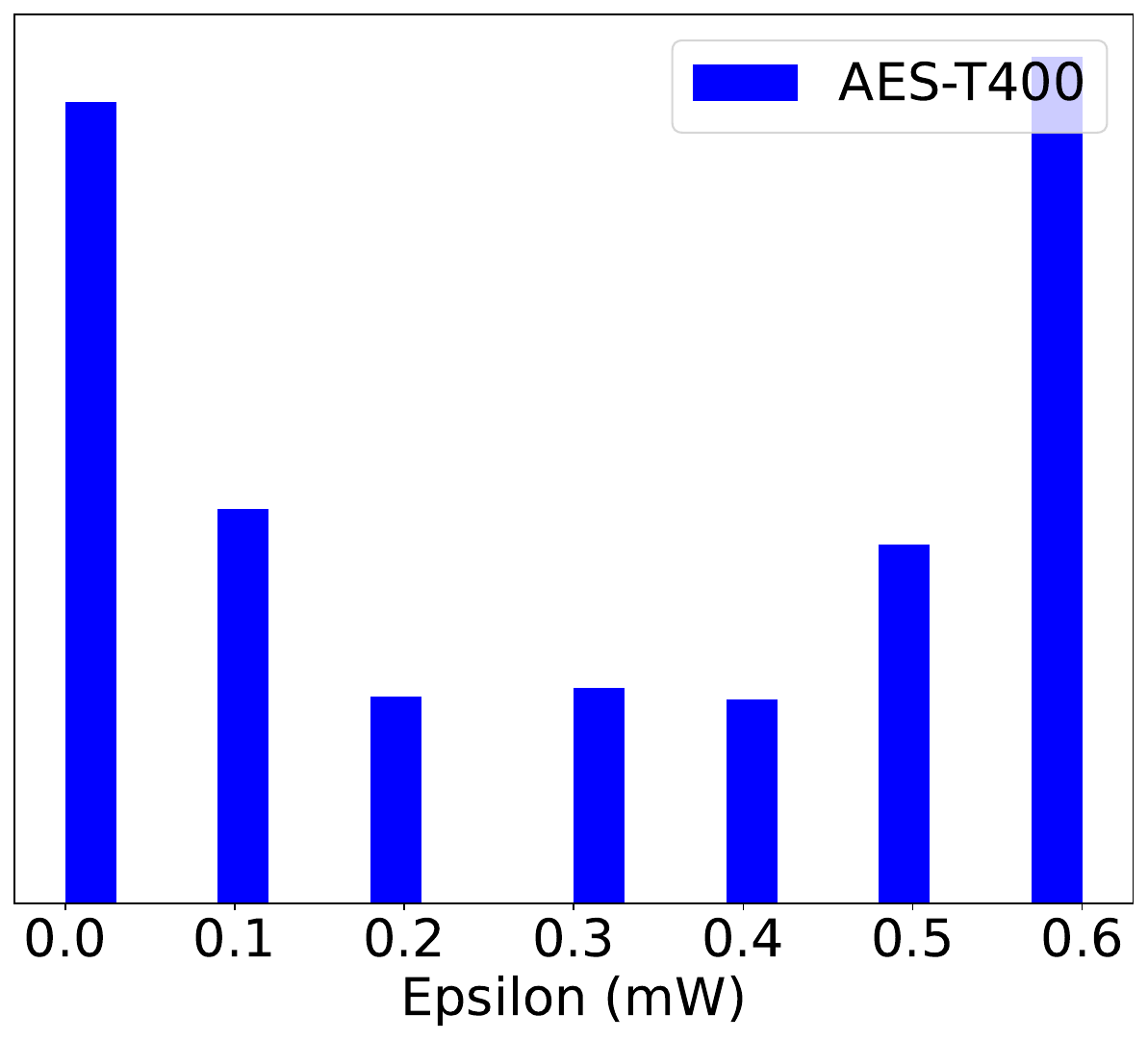}
    	       \caption{AES-T400}
           \end{subfigure}
           
           \begin{subfigure}[b]{\linewidth}
                \includegraphics[width=\linewidth]{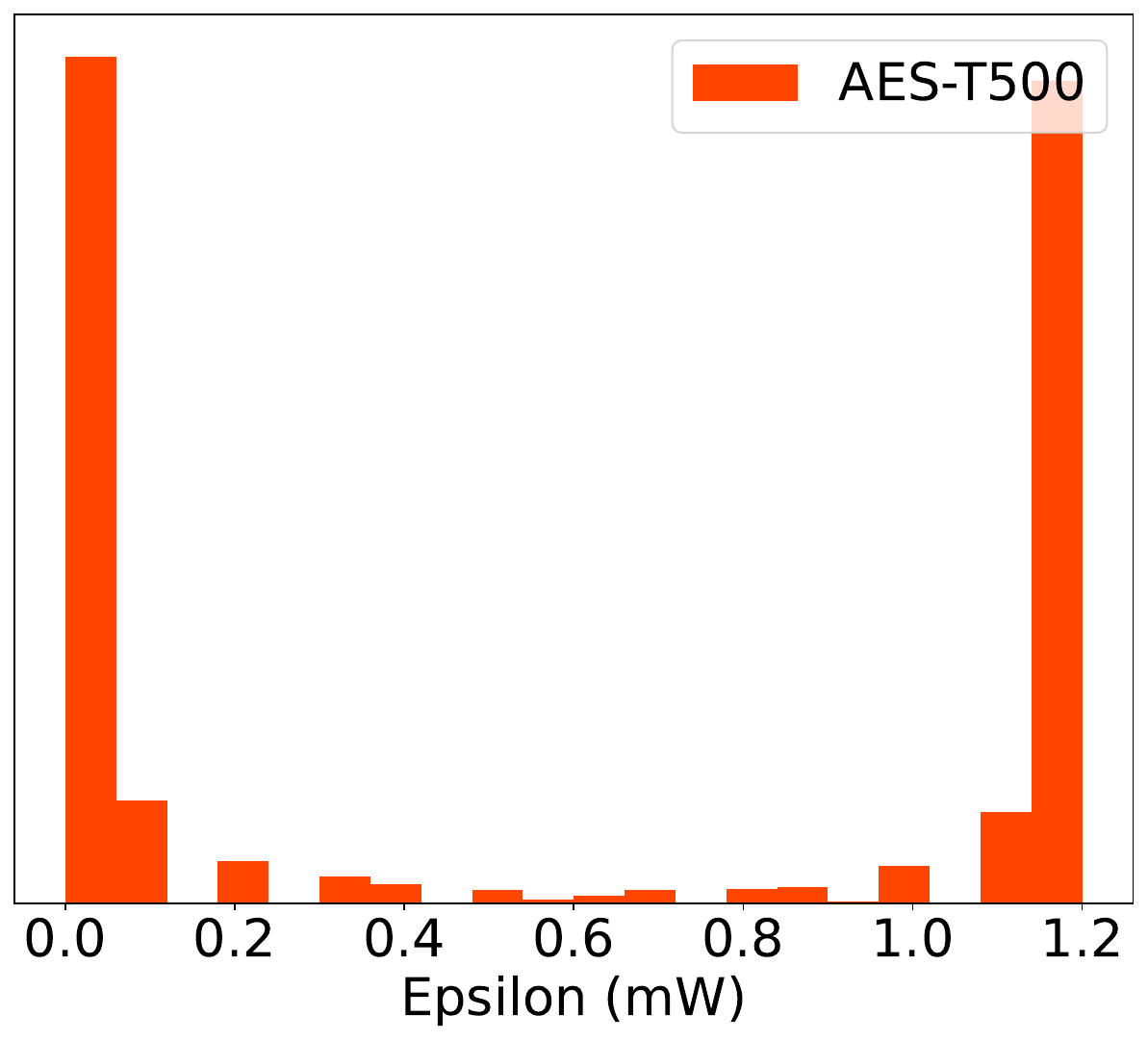}
    	       \caption{AES-T500}
           \end{subfigure}
           
           \begin{subfigure}[b]{\linewidth}
                \includegraphics[width=\linewidth]{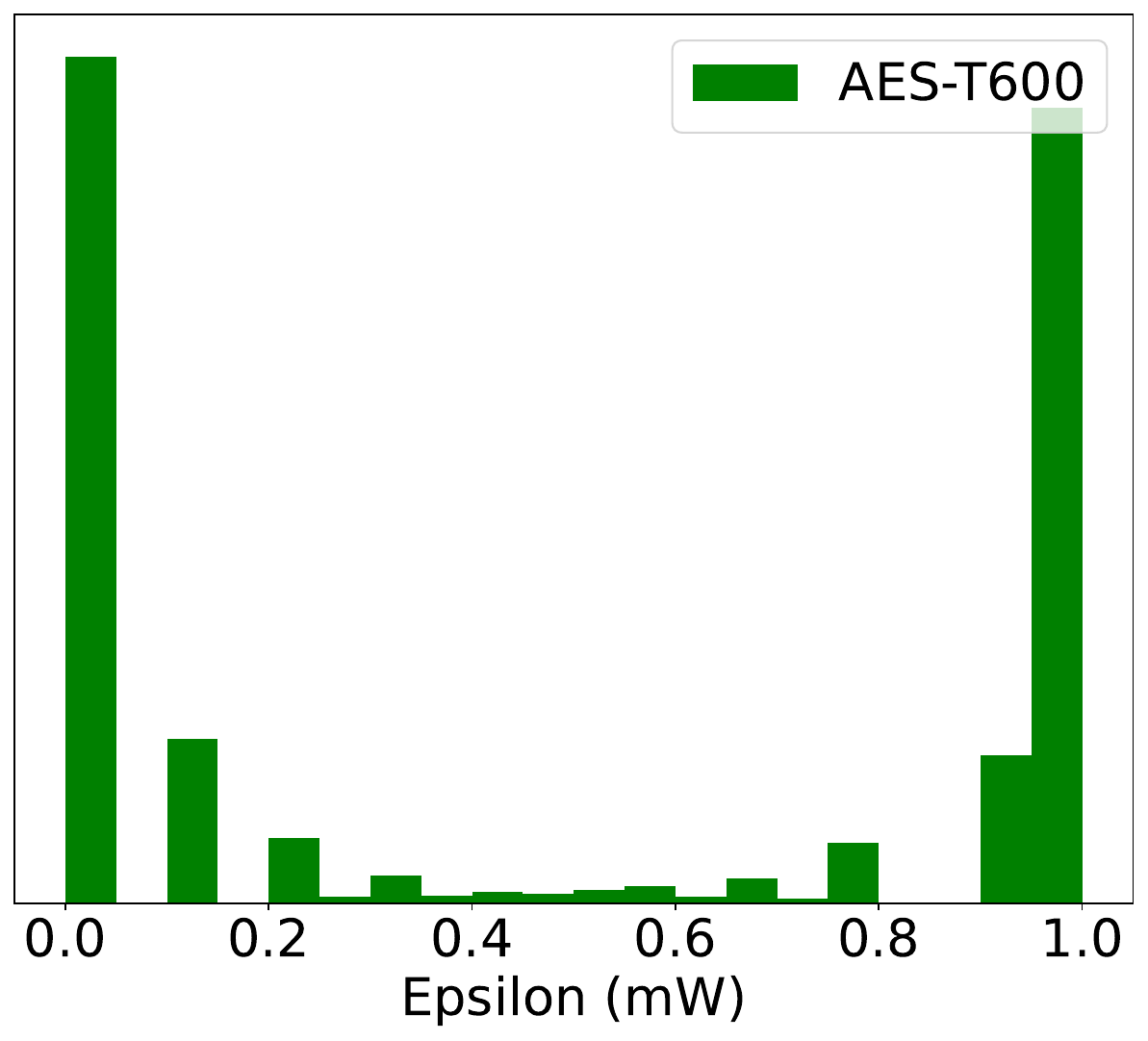}
    	       \caption{AES-T600}
           \end{subfigure}}

           

       \resizebox{\linewidth}{!}{
            \begin{subfigure}[b]{\linewidth}
                \includegraphics[width=\linewidth]{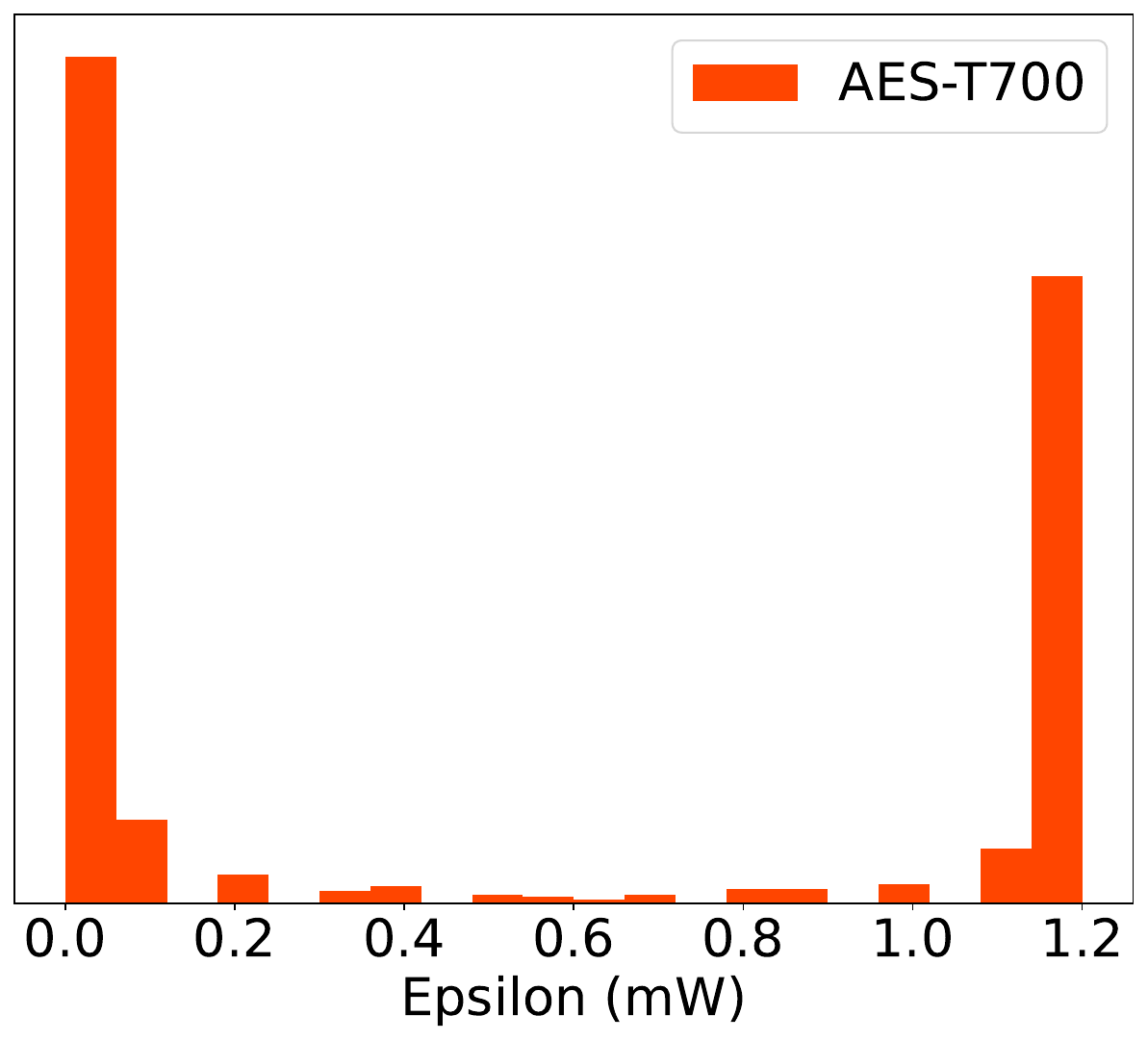}
    	       \caption{AES-T700}
           \end{subfigure}
           
           \begin{subfigure}[b]{\linewidth}
                \includegraphics[width=\linewidth]{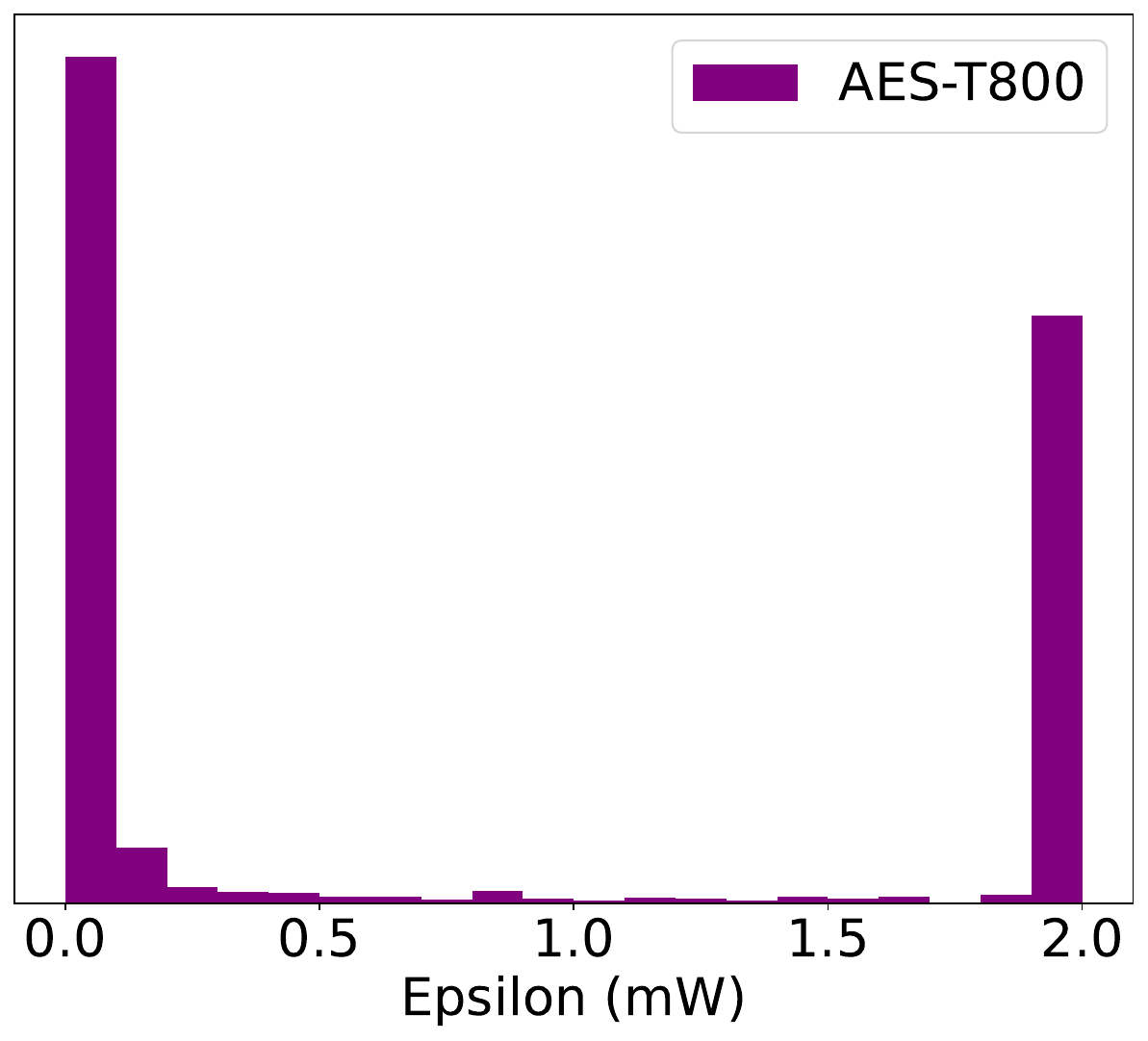}
    	       \caption{AES-T800}
           \end{subfigure}
    
           \begin{subfigure}[b]{\linewidth}
                \includegraphics[width=\linewidth]{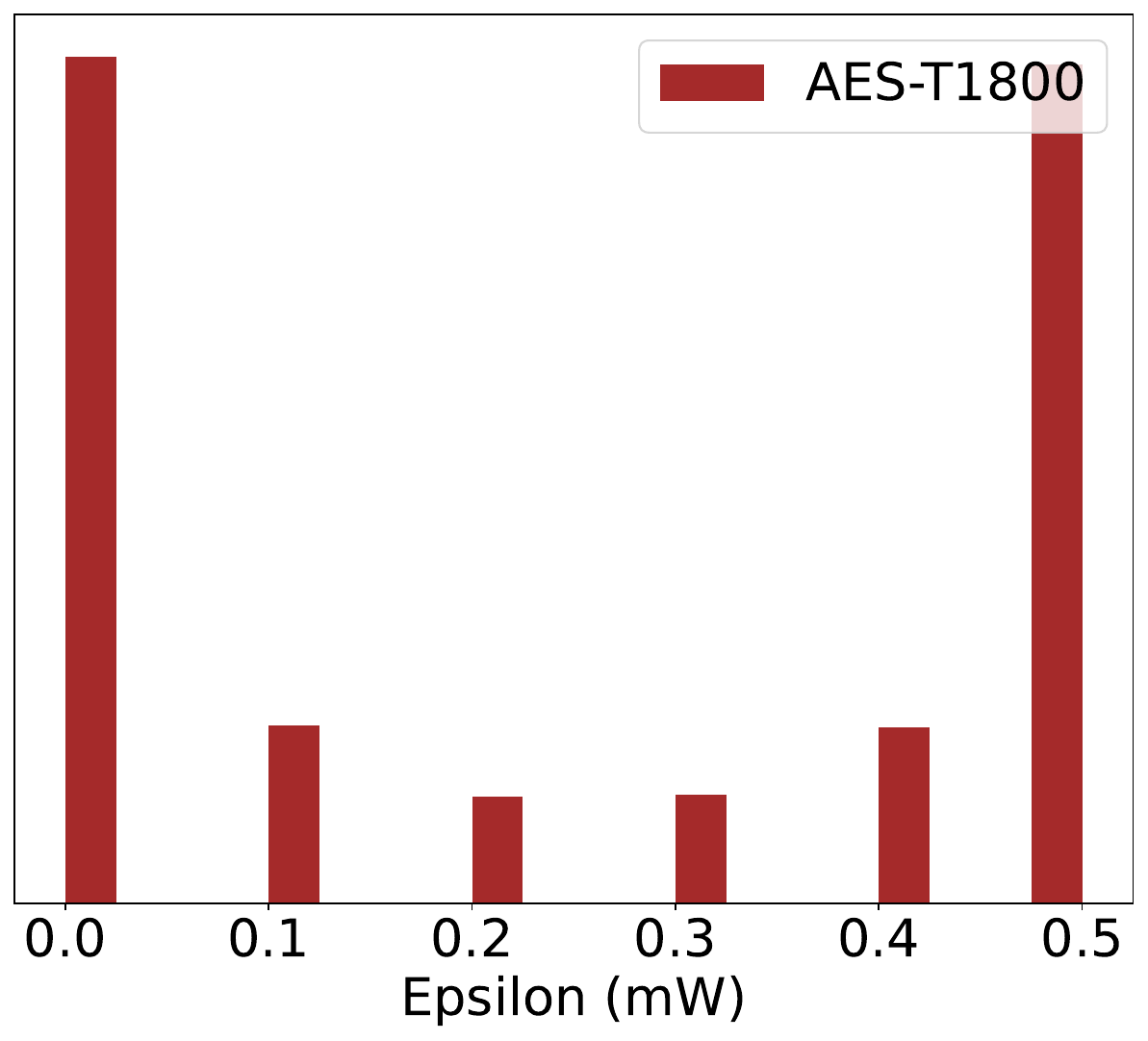}
    	       \caption{AES-T1800}
           \end{subfigure}}

       \caption{Distribution of the adversarial patch values for different target designs.}
	 \label{fig:full_patch_d} 
\end{figure}

Our goal is to study the noise values distribution of the adversarial patch. Our intuition is that if the noise power distribution is biased towards a relatively limited number of values, we can exploit this to optimize the HTO circuit. Figure~\ref{fig:full_patch_d} shows the noise power distribution of a full patch. We can clearly see that in this example the noise power distribution is biased towards $lower$ and $upper$ bound of noise budget.  


\subsection{Adversarial patch quantization}

The previous analysis in Section \ref{sec:patchdist} showed that the noise power distribution is biased towards a relatively limited number of values. Hence, a projection of the adversarial noise to a quantized space representing the most frequent values might maintain its adversarial impact.

To verify and test this hypothesis, we generate new adversarial patches with constraints on their values within a target space. The patch quantization method is exposed in Algorithm \ref{Algo:quantization}. The algorithm updates the adversarial patch generation process by a projection function (Line \rpoint{11}), which takes a vector and rounds its values to the closes values in a subspace $\mathcal{S}$. This subspace is defined according to the most frequent values observed beforehand.

\begin{algorithm}[!ht]
\small
\caption{Quantized Adversarial Power Trace Generation.}
\label{Algo:quantization}
\begin{algorithmic}[1]
\State \textbf{Input:} Data set: $\mathcal{D}$, Subspace: $\mathcal{S}$, classifier: $C$,  noise magnitude: $\varepsilon$, standard deviation: $\sigma$, number of iterations: $N$.

\State \textbf{Output:} $\delta$ (\underline{Quantized} adversarial power trace).\\

\State Initialize $\delta \leftarrow rand()$
\While{$iter < N$ }
    \For{each Batch $X_i \sim \mathcal{D}$}\\
            \State $\delta \leftarrow \frac{1}{|X_i|} \sum_{j=1}^{|X_i|}
             \{ \{\alpha sign(\nabla_{x} J_{\theta}(C({x_{j} + \delta }),\ell))\}\}$ \\
            
            \State $\delta \leftarrow Clip_\varepsilon\{ \delta \} + z \sim N\left(0, \sigma^{2}\right)$
             \State $\delta \leftarrow \mathcal{P}_{\mathcal{S}}(\delta)$
    \EndFor
\State $iter += 1$
\EndWhile

\end{algorithmic}
\end{algorithm}


\subsection{Results} 

\begin{figure}[!ht]
    \centering
	\includegraphics[width=0.7\columnwidth]{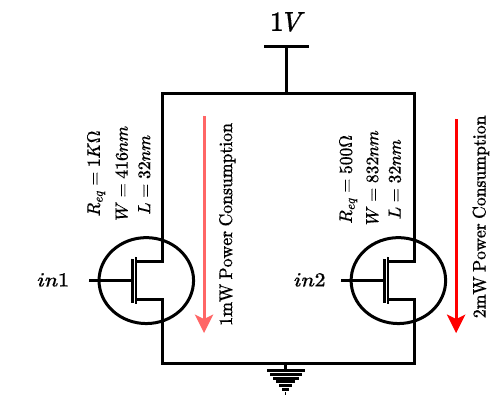}
	\caption{Optimized ASIC adversarial noise generator }
	\label{fig:optimized_power_network_3_1} 
\end{figure}

Generating patches with different resolutions ($0.1 mW$ to $1 mW$) lead to a significant discovery that still with the resolution of $1 mW$ we reach $100\%$ efficiency in misclassifying the ML-based Trojan detector. Therefore, to reduce the resource utilization of the adversarial power generator circuit for a more sneaky attack, we put the equivalent transistor consuming exactly the power in the noise patch. Figure \ref{fig:optimized_power_network_3_1} depicts an optimized circuit that can consume power in the range $0 mW$ to $2 mW$ with the resolution of $1 mW$ with only $2$ transistors.

Furthermore, we leverage different optimization processes on patches to reduce the resource utilization on ASICs and FPGAs:

\begin{itemize}
    \item Calculating the mean value of adversarial power trace and project other samples to the nearest value to the upper bound, lower bound, and mean value (2 transistors)
    \item Mapping above the mean value to the upper bound and below ones to zero (1 transistor) 
\end{itemize}

Exploiting various quantization spaces in adversarial patches decreases the number of transistors used in the design. Therefore, it minimizes the cost and area of producing the HTO in the ASIC platform. Moreover, this optimization lowers the configuration vectors causing a significant reduction in storage units for both platforms. Our results demonstrate that we still can keep the accuracy of the model at the acceptable value by quantizing the values of patches to $2$ or even $1$. Table \ref{table:optimized_ASIC_synchronized} and \ref{table:optimized_FPGA_synchronized} represent resource utilization for optimized patches on ASICs and FPGA respectively.




\begin{table}[!ht]
\small
    \centering
    \resizebox{\linewidth}{!}{
    \begin{tabular}{| C{1.7cm} | C{2cm} | C{1.5cm}| C{2cm} | C{1.5cm}|}
             \hline
             \multirow{3}{1.7cm}{\centering Dataset}&\multirow{3}{2cm}{\centering HT detection accuracy (\%)} &\multirow{3}{1.5cm}{\centering \# of transistors} &  \multirow{3}{2cm}{ \centering HT detection accuracy after patching (\%)} & \multirow{3}{1.5cm}{\centering Noise Budget (mW)}\\
             &&&&\\
             &&&& \\\hline\hline

             \multirow{1}{1.7cm}{\centering AES-T400} & \multirow{1}{2cm}{\centering 97.1} & 1  & 0 & \multirow{1}{1.5cm}{\centering 0-0.6} \\\hline

             \multirow{1}{1.7cm}{\centering AES-T500} & \multirow{1}{2cm}{\centering 91.91} & 1  & 0 & \multirow{1}{1.5cm}{\centering 0-1.2}\\\hline

             \multirow{1}{1.7cm}{\centering AES-T600} & \multirow{1}{2cm}{\centering 93.98} & 1 & 0 & \multirow{1}{1.5cm}{\centering 0-1} \\\hline

             \multirow{1}{1.7cm}{\centering AES-T700} & \multirow{1}{2cm}{\centering 100} & 1 & 0 & \multirow{1}{1.5cm}{\centering 0-1.2} \\\hline

             \multirow{1}{1.7cm}{\centering AES-T800} & \multirow{1}{2cm}{\centering 100} & 1 & 0 & \multirow{1}{1.5cm}{\centering 0-2}\\\hline

             \multirow{1}{1.7cm}{\centering AES-T1800} & \multirow{1}{2cm}{\centering 94.12} & 1  & 0 & \multirow{1}{1.5cm}{\centering 0-0.5}\\\hline              
    \end{tabular}
    }
    
    \caption{Optimized resource utilization and HT detection accuracy on ASIC}
    \label{table:optimized_ASIC_synchronized}
\end{table}

\begin{table}[!ht]
\small
    \centering
    \resizebox{\linewidth}{!}{
    \begin{tabular}{| C{1.7cm} | C{2cm} | C{1cm} | C{1.5cm}| C{2cm}   | C{1.5cm}|}
             \hline
             \multirow{3}{1.7cm}{\centering Dataset} & \multirow{3}{2cm}{\centering HT detection accuracy (\%)} & \multirow{3}{1cm}{\centering Method} & \multirow{3}{1.5cm}{\centering \# of CLBs} & \multirow{3}{2cm}{ \centering HT detection accuracy after patching (\%)} & \multirow{3}{1.5cm}{\centering Noise Budget (mW)}\\
             &&&&&\\
             &&&&&\\\hline\hline

             \multirow{2}{1.7cm}{\centering AES-T400} & \multirow{2}{2cm}{\centering 97.1} & RO & 2 &  0 & \multirow{2}{1.5cm}{\centering 0-1} \\\cline{3-5}
             && DSP & 1  & 0 & \\\hline

             \multirow{2}{1.7cm}{\centering AES-T500} & \multirow{2}{2cm}{\centering 91.91} & RO & 4  & 0 & \multirow{2}{1.5cm}{\centering 0-2} \\\cline{3-5}
             && DSP & 2 & 0 & \\\hline

             \multirow{2}{1.7cm}{\centering AES-T600} & \multirow{2}{2cm}{\centering 93.84} & RO & 2  & 0.7 & \multirow{2}{1.5cm}{\centering 0-1} \\\cline{3-5}
             && DSP & 1  & 0 & \\\hline

             \multirow{2}{1.7cm}{\centering AES-T700} & \multirow{2}{2cm}{\centering 100} & RO & 4  & 0.2 & \multirow{2}{1.5cm}{\centering 0-2} \\\cline{3-5}
             && DSP & 2  & 0 & \\\hline

             \multirow{2}{1.7cm}{\centering AES-T800} & \multirow{2}{2cm}{\centering 100} & RO & 4  & 0 & \multirow{2}{1.5cm}{\centering 0-2} \\\cline{3-5}
             && DSP & 2  & 0 & \\\hline

             \multirow{2}{1.7cm}{\centering AES-T1800} & \multirow{2}{2cm}{\centering 94.12} & RO & 2  & 0 & \multirow{2}{1.5cm}{\centering 0-1} \\\cline{3-5}
             && DSP & 1  & 0 & \\\hline               
    \end{tabular}
    }
    
    \caption{Optimized resource utilization and HT detection accuracy on FPGA}
    \label{table:optimized_FPGA_synchronized}
\end{table}


\section{Unsynchronized Adversarial Patch}
So far, we assumed that the power trace and the adversarial noise are synchronised, which can be implemented by a simple triggering signal, e.g., the HTO circuit is triggered by first round of the AES. In the section, we relax this assumption, and suppose the starting point of capturing power traces can change and drop the synchronisation assumption. We propose to generate a patch that is robust in this setting. Taking into account this characteristic of the attack, we propose a patch generator that considers the shift existing in the setup. Hence, the new problem can be formulated as follows:
\begin{equation}
\small
\begin{split}
        C(x + shift(\delta, k))) \neq C(x), ~ \forall ~ x \sim \mu_{HT}, and \\
        \left\Vert \delta \right\Vert_\infty \leq \varepsilon \\
\end{split}
\end{equation}

Where $\varepsilon$ controls the magnitude of the noise vector ($\delta$), $shift(.)$ is a function that quantifies the adversarial patch $\delta$ relatively with regard to a target power trace $y$ given a time shift $k \in [0, d] $. Given an adversarial patch  $\delta=\{\delta_j\} \forall j \in[0, d] $ that is \emph{consumed by a circuit in a continuous loop}, the function $shift(.)$ could be expressed as:

\begin{equation}\label{eq:shift}
\small
\begin{split}
shift(\delta, k)=\left\{\begin{array}{lll}
\delta(d-k+j) & \text{if} & j \in[0, k] \\
\delta(j-k) & ~ \text{else}
\end{array}\right.\\
\end{split}
\end{equation}

To solve the problem formulated above, we propose Algorithm \ref{Algo:unsynch}, which details the noise generation mechanism of an HTO-generated power trace that is unsynchronised with the victim circuit.

\begin{algorithm}[!ht]
\small
\caption{Un-synchronised Adversarial Power Trace Generation.}
\label{Algo:unsynch}
\begin{algorithmic}[1]
\State \textbf{Input:} Data set: $\mathcal{D}$, classifier: $C$,  noise magnitude: $\varepsilon$, standard deviation: $\sigma$, number of iterations: $N$.

\State \textbf{Output:} $\delta$ Adversarial power trace.\\

\State Initialize $\delta \leftarrow rand()$
\While{$iter < N$ }
    \For{each Batch $X_i \sim \mathcal{D}$}\\
            
            \State $\delta \leftarrow \frac{1}{|X_i|} \sum_{j=1}^{|X_i|}
             \{ \{\alpha sign(\nabla_{x} J_{\theta}(C({x_{j} + sh(\delta) }),\ell))\}\}$ \\
            
            \State $\delta \leftarrow Clip_\varepsilon\{ \delta \} + z \sim N\left(0, \sigma^{2}\right)$
    \EndFor
\State $iter += 1$
\EndWhile\\

\Function{$sh$}{$\delta$}\Comment{ \textcolor{gray}{//Time shift function.}}

\State $k \leftarrow Rand(d-1)$ \Comment{ \textcolor{gray}{//Select a random shift.}}
\State Initialize $shifted\_\delta \leftarrow 0$
\State $shifted\_\delta \leftarrow concat(\delta [d-k:d],\delta [0:d-k] )$
\State \textbf{Return:}$shifted\_\delta$
\EndFunction

\end{algorithmic}
\end{algorithm}

\begin{figure}[!ht]
    \centering
	\resizebox{\linewidth}{!}{\includegraphics{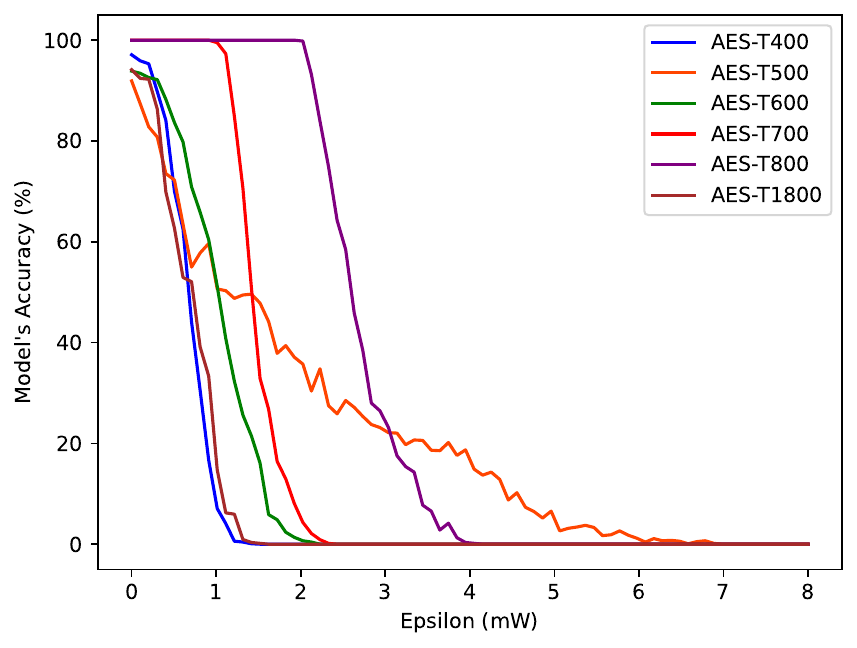}}
	\caption{HT detection accuracy based on changing the maximum power budget for unsynchronized patches. The stochasticity comes from the random time shift applied to the patch at inference.}
       \label{fig:model_accuracy_for_unsynch} 
\end{figure}

The effectiveness of different power budgets on model accuracy for the unsynchronized patches has been evaluated in Figure \ref{fig:model_accuracy_for_unsynch}.
The figure shows that for fooling the ML model with $100\%$
efficiency we can restrict the maximum power of adversarial
patches to $6.9mW$.

For unsynchronized patches, which consume maximum power in the range $0mW$ to $6.9mW$ ($AES-T500$), the mean square errors of $0.025$ and $0.21$ have been calculated for ASIC and FPGA respectively. The maximum exploited resources for the simulation of unsynchronized patches with the resolution of $1mW$ are $14$ ring-oscillators in FPGA. On the other hand, to simulate patches with the resolution of $0.1mW$, we require the maximum of $16$ transistors.



\begin{figure}[!ht]
\centering
        \captionsetup[sub]{font=huge}
	\resizebox{0.95\linewidth}{!}{
 \centering
        \begin{subfigure}[b]{\linewidth}
            \includegraphics[width=\linewidth]{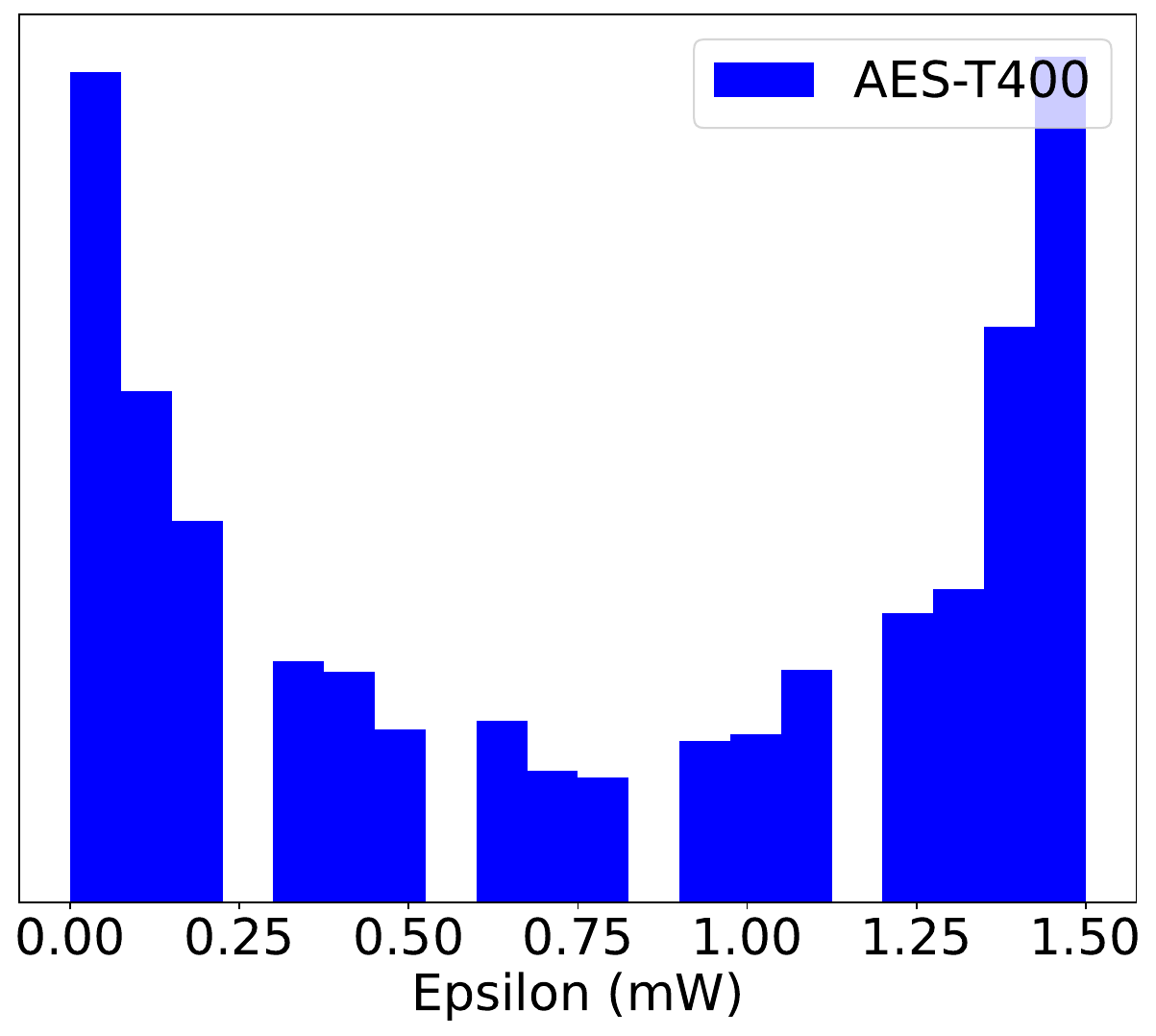}
	       \caption{AES-T400}
       \end{subfigure}
       
       \begin{subfigure}[b]{\linewidth}
            \includegraphics[width=\linewidth]{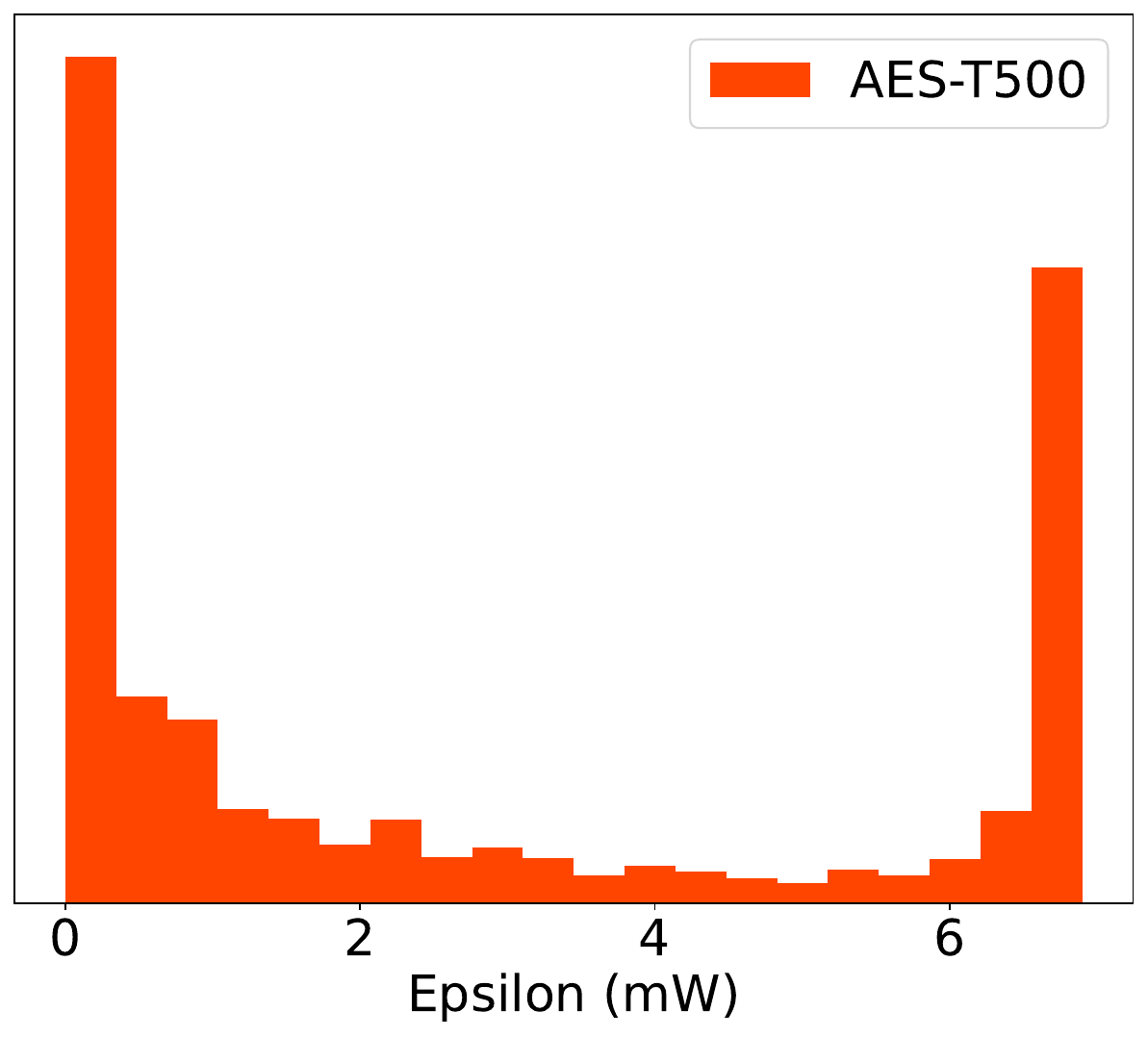}
	       \caption{AES-T500}
       \end{subfigure}

       \begin{subfigure}[b]{\linewidth}
            \includegraphics[width=\linewidth]{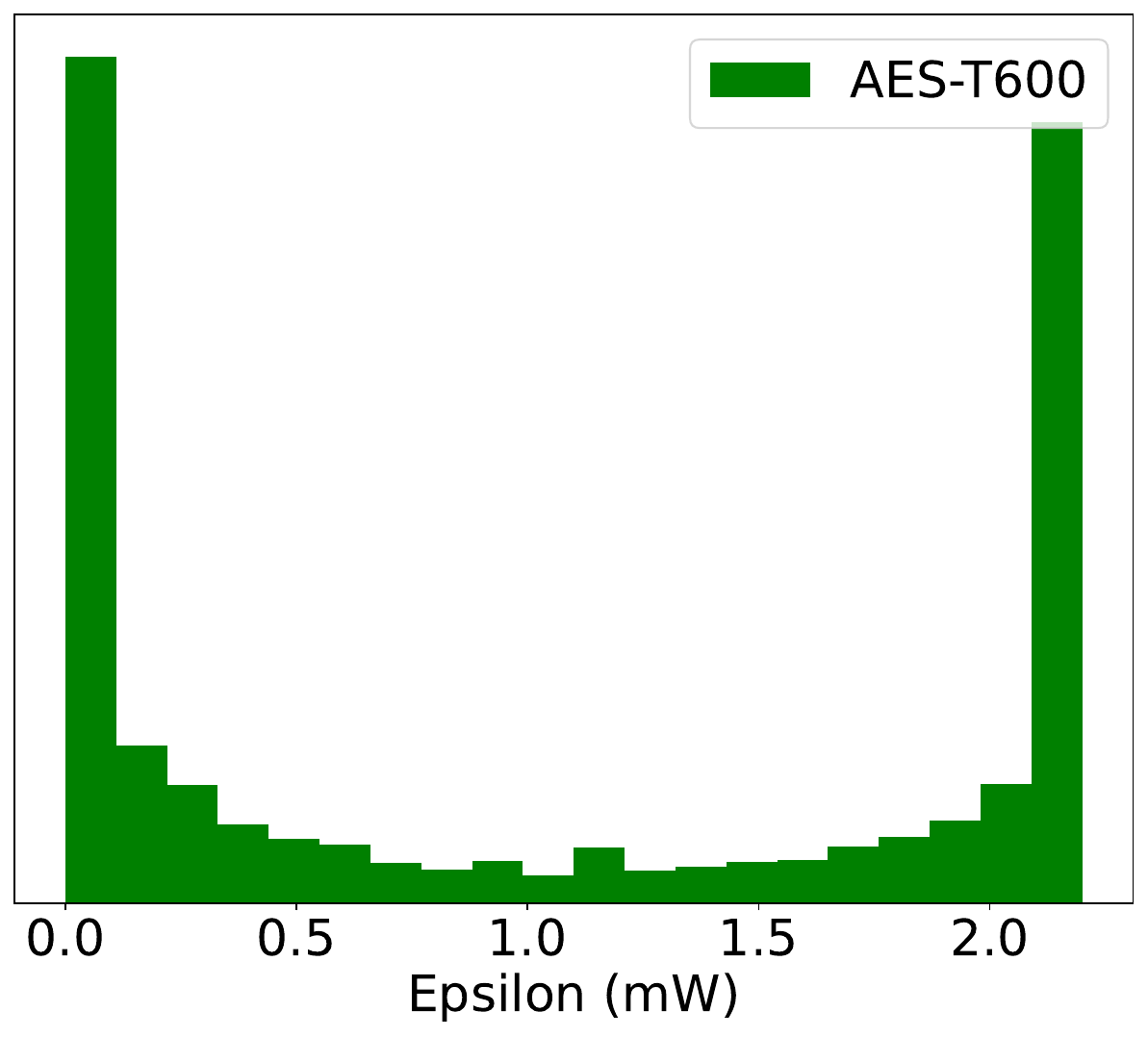}
	       \caption{AES-T600}
       \end{subfigure}}

       \resizebox{0.95\linewidth}{!}{
        \centering
        \begin{subfigure}[b]{\linewidth}
            \includegraphics[width=\linewidth]{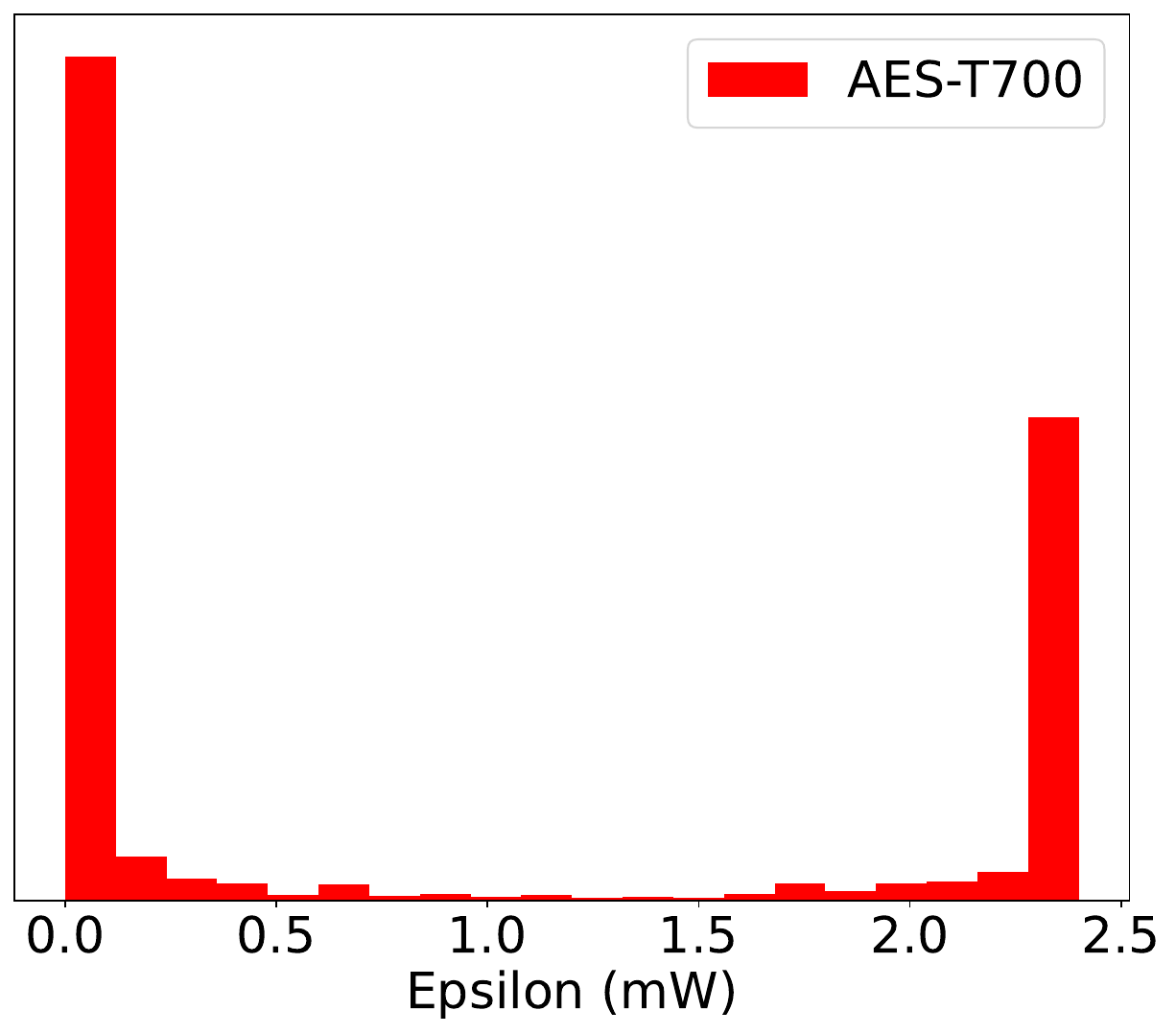}
	       \caption{AES-T700}
       \end{subfigure}
       
       \begin{subfigure}[b]{\linewidth}
            \includegraphics[width=\linewidth]{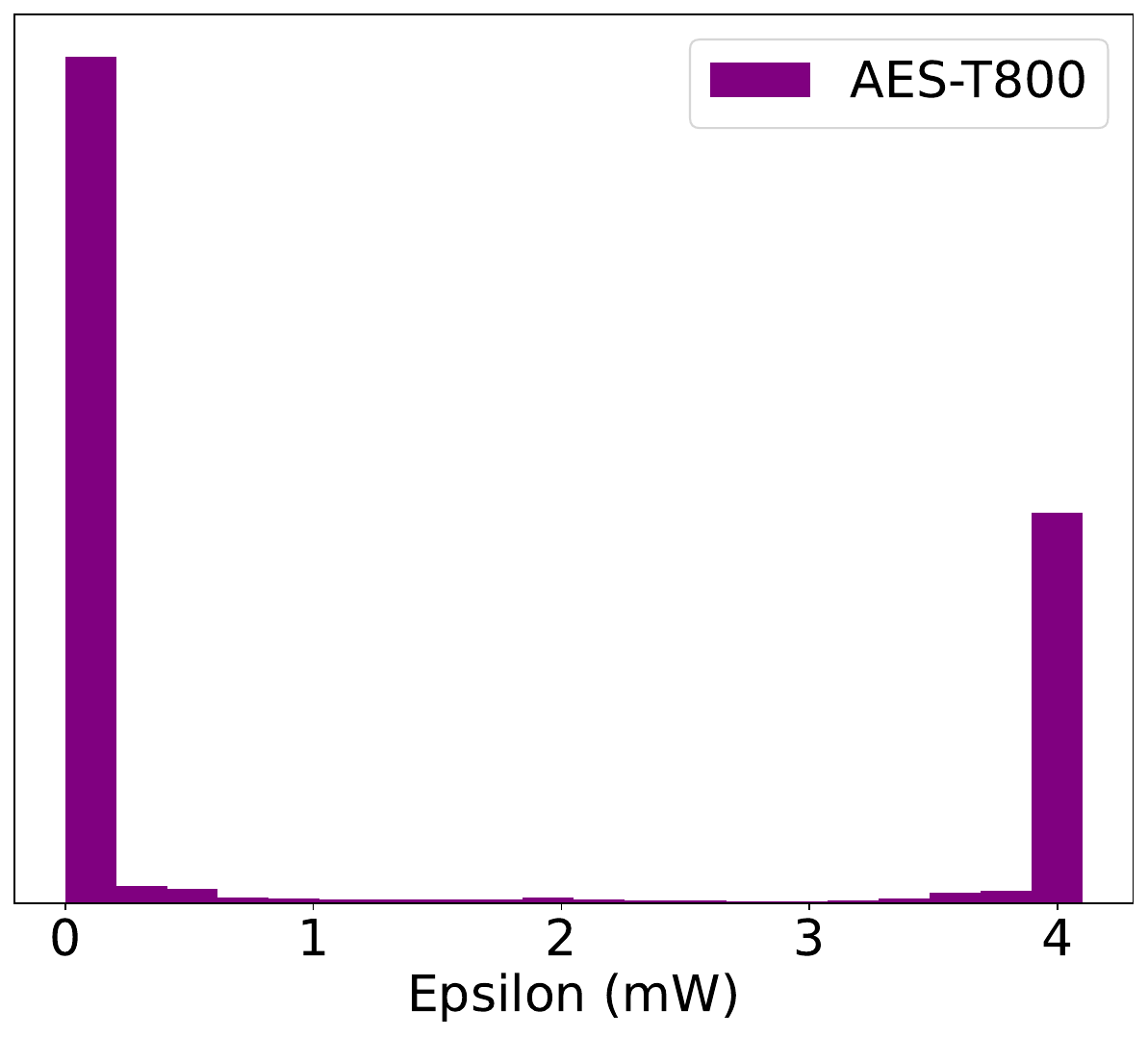}
	       \caption{AES-T800}
       \end{subfigure}

       \begin{subfigure}[b]{\linewidth}
            \includegraphics[width=\linewidth]{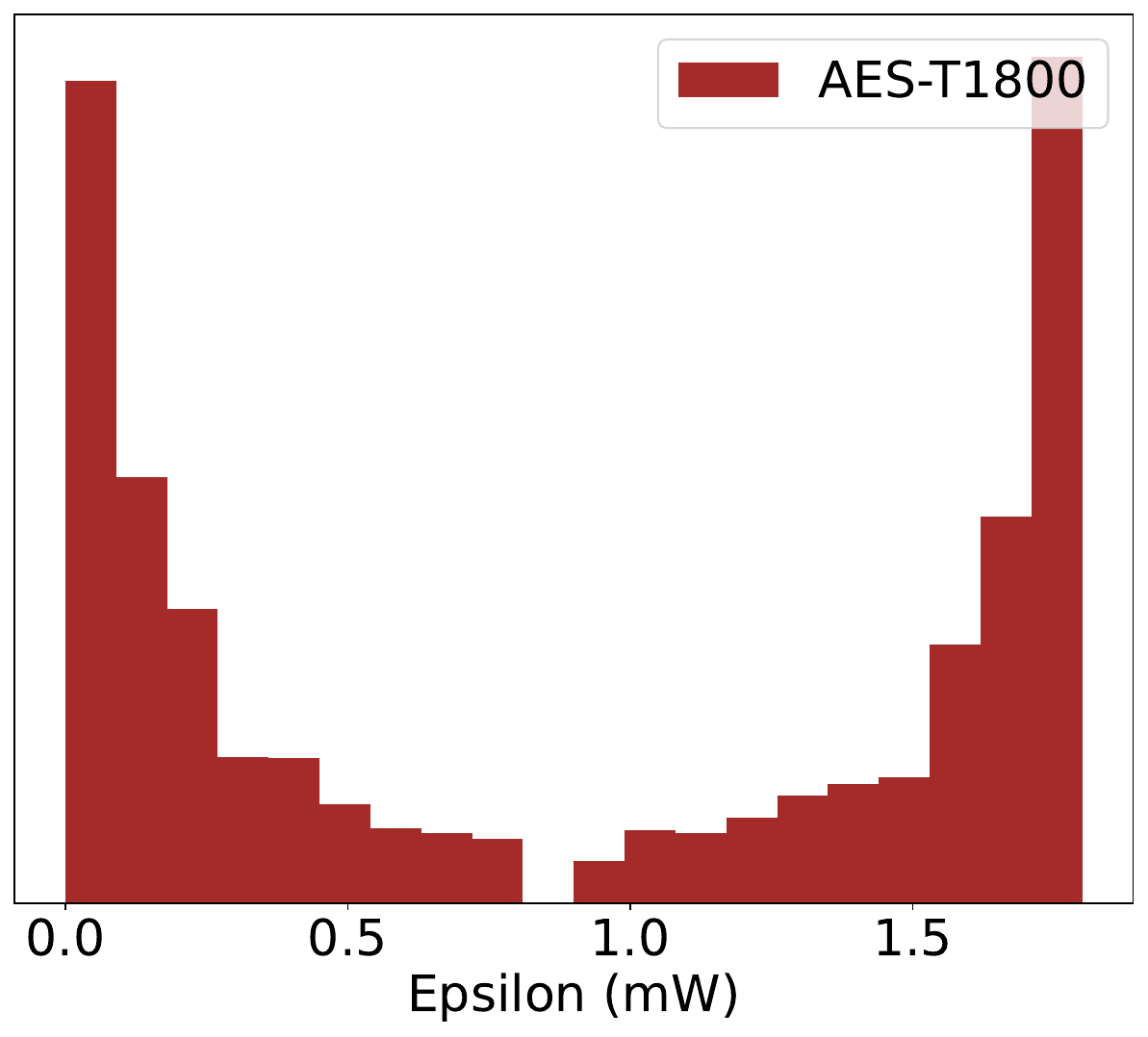}
	       \caption{AES-T1800}
       \end{subfigure}}

       \caption{Full unsynchronized patch distribution}
	 \label{fig:unsync_full_patch_d} 
\end{figure}

Interestingly, similar to patches generated based on Algorithm \ref{Algo:patchGen}, as illustrated in Figure \ref{fig:unsync_full_patch_d}, the noise power distribution is restricted to a relatively limited number of values. Hence, we leverage the same scenario discussed in Section \ref{sec:HTO_opt} to optimize the circuits for unsynchronized patches. We use the same quantisation methodology and reduce the values of patches to a subspace that still enables effective adversarial patches implementable with lower resources.
The resource utilization and model's accuracy of the unsynchronized patches are shown in Tables \ref{table:ASIC_unsynchronized} and \ref{table:FPGA_unsynchronized}

\begin{table}[!ht]
\small
    \centering
    \resizebox{\linewidth}{!}{
    \begin{tabular}{| C{1.7cm} | C{2cm} | C{1.5cm}| C{2cm} | C{1.5cm}|}
             \hline
             \multirow{3}{1.7cm}{\centering Dataset}&\multirow{3}{2cm}{\centering HT detection accuracy (\%)} &\multirow{3}{1.5cm}{\centering \# of transistors} &  \multirow{3}{2cm}{ \centering HT detection accuracy after patching (\%)} & \multirow{3}{1.5cm}{\centering Noise Budget (mW)}\\
             &&&&\\
             &&&& \\\hline\hline

             \multirow{2}{1.7cm}{\centering AES-T400} & \multirow{2}{2cm}{\centering 97.1} & 11 &  0 & \multirow{2}{1.5cm}{\centering 0-1.5} \\\cline{3-4}
             && 1 &  0 & \\\hline

             \multirow{2}{1.7cm}{\centering AES-T500} & \multirow{2}{2cm}{\centering 91.91} & 16  & 0 & \multirow{2}{1.5cm}{\centering 0-6.9} \\\cline{3-4}
             && 1 &  0 & \\\hline

             \multirow{2}{1.7cm}{\centering AES-T600} & \multirow{2}{2cm}{\centering 93.87} & 12  & 0 & \multirow{2}{1.5cm}{\centering 0-2.2} \\\cline{3-4}
             && 1  & 0 & \\\hline

             \multirow{2}{1.7cm}{\centering AES-T700} & \multirow{2}{2cm}{\centering 100} & 12 & 0 & \multirow{2}{1.5cm}{\centering 0-2.4} \\\cline{3-4}
             && 1 &  0 & \\\hline

             \multirow{2}{1.7cm}{\centering AES-T800} & \multirow{2}{2cm}{\centering 100} & 14  & 0 & \multirow{2}{1.5cm}{\centering 0-4.1} \\\cline{3-4}
             && 1 &  0.1 & \\\hline

             \multirow{2}{1.7cm}{\centering AES-T1800} & \multirow{2}{2cm}{\centering 94.12} & 11  & 0 & \multirow{2}{1.5cm}{\centering 0-1.6} \\\cline{3-4}
             && 1 &  0 & \\\hline              
    \end{tabular}
    }
    
    \caption{Resource utilization and HT detection accuracy on ASIC for unsynchronized patches}
    \label{table:ASIC_unsynchronized}
\end{table}

\begin{table}[!ht]
\small
    \centering
    \resizebox{\linewidth}{!}{
    \begin{tabular}{| C{1.7cm} | C{2cm} | C{1cm} | C{1.5cm} | C{2cm}  | C{1.5cm}|}
             \hline
             \multirow{3}{1.7cm}{\centering Dataset} & \multirow{3}{2cm}{\centering HT detection accuracy (\%)} & \multirow{3}{1cm}{\centering Method} & \multirow{3}{1.5cm}{\centering \# of CLBs} & \multirow{3}{2cm}{ \centering HT detection accuracy after patching (\%)} & \multirow{2}{1.5cm}{\centering Noise Budget (mW)}\\
             &&&&&\\
             &&&&& \\\hline\hline

             \multirow{2}{1.7cm}{\centering AES-T400} & \multirow{2}{2cm}{\centering 97.1} & RO & 4  & 0 & \multirow{2}{1.5cm}{\centering 0-2} \\\cline{3-5}
             && DSP & 2 & 0 & \\\hline

             \multirow{2}{1.7cm}{\centering AES-T500} & \multirow{2}{2cm}{\centering 91.91} & RO & 14  & 0 & \multirow{2}{1.5cm}{\centering 0-7} \\\cline{3-5}
             && DSP & 7  & 0 & \\\hline

             \multirow{2}{1.7cm}{\centering AES-T600} & \multirow{2}{2cm}{\centering 93.87} & RO & 6  & 0 & \multirow{2}{1.5cm}{\centering 0-3} \\\cline{3-5}
             && DSP & 3  & 0 & \\\hline

             \multirow{2}{1.7cm}{\centering AES-T700} & \multirow{2}{2cm}{\centering 100} & RO & 6  & 0.1 & \multirow{2}{1cm}{\centering 0-3} \\\cline{3-5}
             && DSP & 3 &  0 & \\\hline

             \multirow{2}{1.7cm}{\centering AES-T800} & \multirow{2}{2cm}{\centering 100} & RO & 8 & 0.3 & \multirow{2}{1.5cm}{\centering 0-4} \\\cline{3-5}
             && DSP & 4 & 0 & \\\hline

             \multirow{2}{1.7cm}{\centering AES-T1800} & \multirow{2}{2cm}{\centering 94.12} & RO & 4  & 0 & \multirow{2}{1.5cm}{\centering 0-2} \\\cline{3-5}
             && DSP & 2 &  0 & \\\hline               
    \end{tabular}
    }
    
    \caption{Resource utilization and HT detection accuracy on FPGA for unsynchronized patches}
    \label{table:FPGA_unsynchronized}
\end{table}


\section{Countermeasures}\label{sec:counter}

 Our initial threat model assumes a defender deploying an ML model for HT detection without any further specific countermeasures. One of the possible protections for such systems is to pre-process the input power trace to filter out undesired signals from non-relevant frequency ranges. In the following, we perform spectral analysis to check if the adversarially generated power trace has a significantly different spectral signature compared to any circuit (be it benign or an HT). Based on this spectral analysis, we propose a pre-processing technique and discuss its limits later. Furthermore, we adversarially trained the HT detector and show that it resulted in a loss in accuracy, while allowing adaptive attacks to bypass the detection with a higher amount of power budget. 

\noindent
\textit{\textbf{Spectral Analysis:}}
The power trace can be represented in the spectral domain using the Fast Fourier Transform (FFT), which is defined by
\begin{equation}
\small
    \mathcal{S}(k) = \sum^{N-1}_{n=0}s(n)e^{-i\frac{2\pi}{N}nk}
\end{equation}


Where $N$ is the length of the spectral signature, $\mathcal{S}(k)$ is the frequency domain magnitudes and $s(n)$ is the time domain power trace in our case.

We consider a system that is comprehensively pre-processing the input signal. The defender defines its target frequency range based on the expected power traces. We evaluate the efficiency of the HTO under this setting. We assume the defender has access to significant potential HTs. We first explore the power spectral density of the benign and HT circuits to identify the frequency range that represents the region of interest ($[f_{min}, f_{max}]$). A comprehensive band-pass filter corresponding to this frequency range is then applied to the collected power traces before feeding it to the ML model.

\begin{figure*}[!ht]
\centering
        \captionsetup[sub]{font=Huge}
	\resizebox{0.95\textwidth}{!}{
            \begin{subfigure}[b]{\textwidth}
                \includegraphics[width=\textwidth]{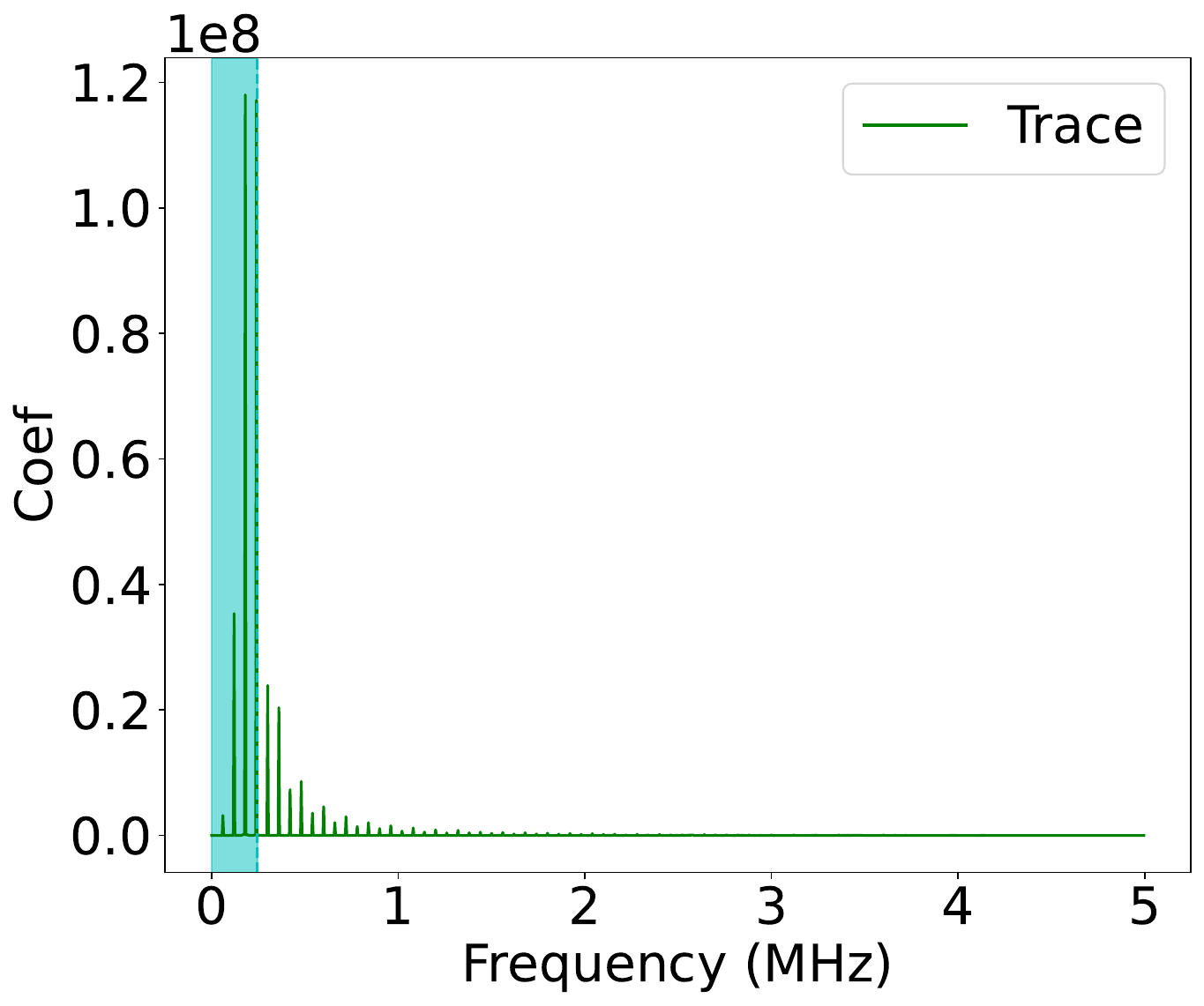}
    	       \caption{FFT(trace)}
            
           \end{subfigure}
           
           \begin{subfigure}[b]{\textwidth}
                \includegraphics[width=\textwidth]{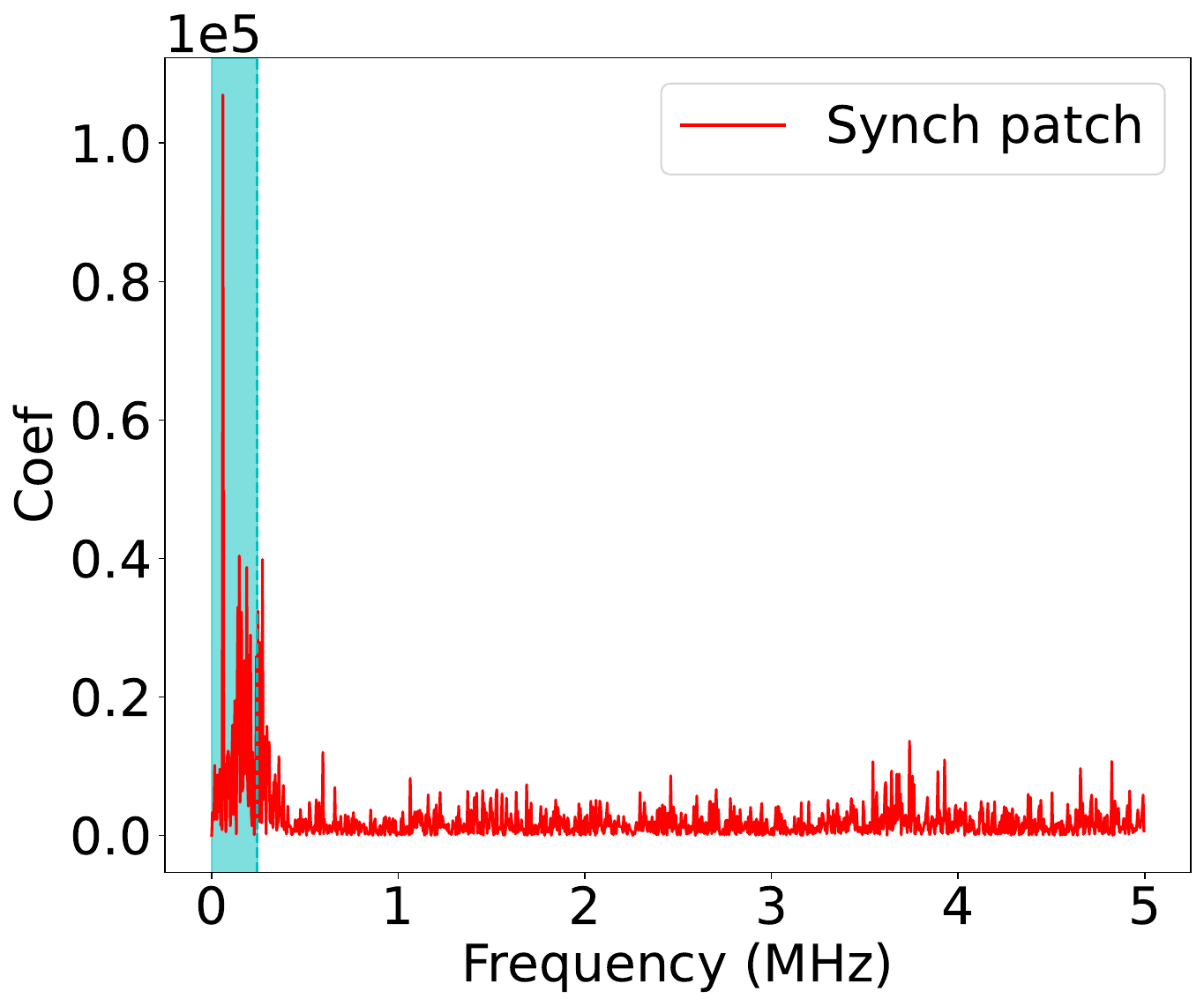}
    	       \caption{FFT(sync patch)}
           \end{subfigure}
    
           \begin{subfigure}[b]{\textwidth}
                \includegraphics[width=\textwidth]{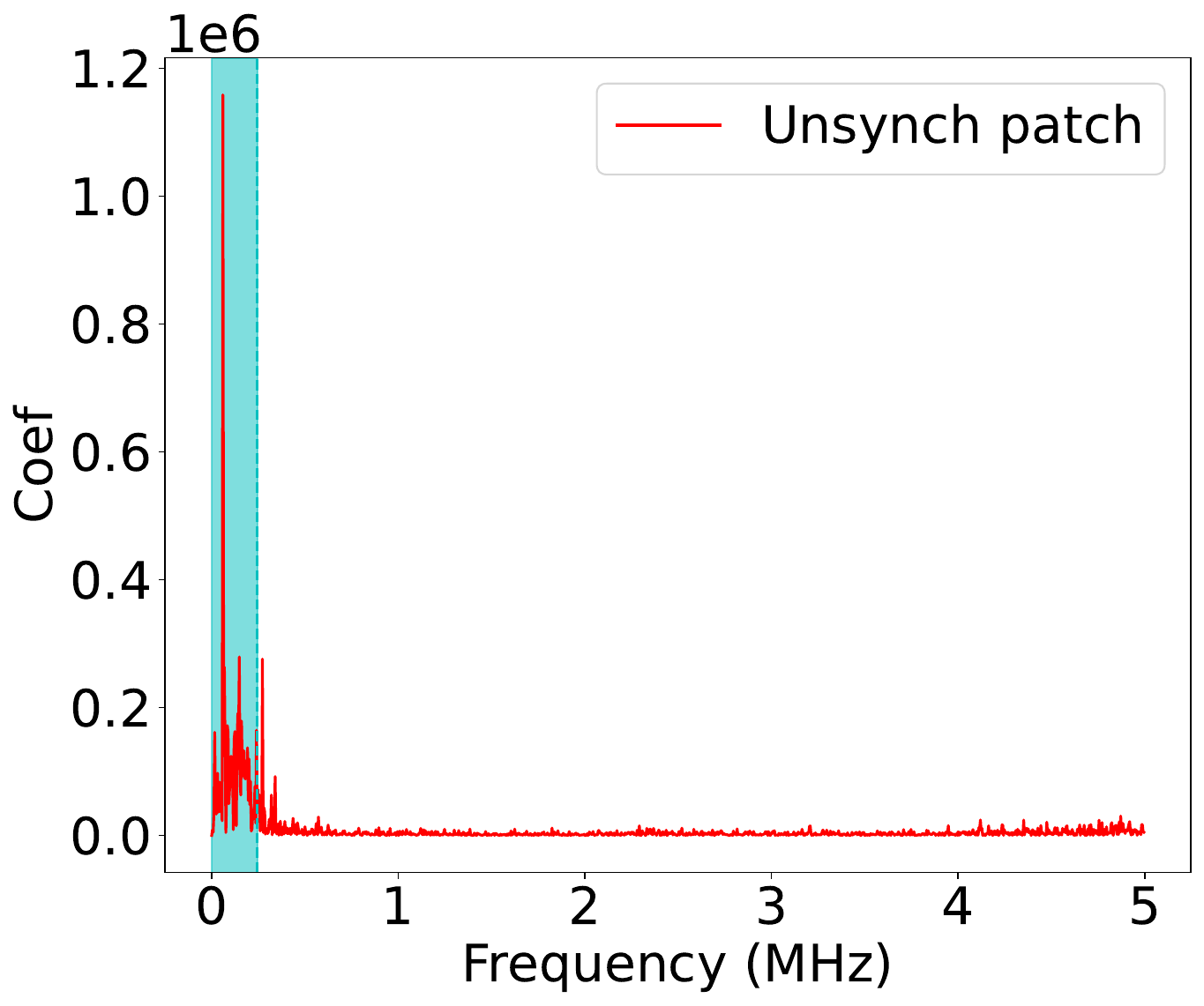}
    	       \caption{FFT(unsync patch)}
           \end{subfigure}}
           
    \caption{AES-T700 and patch power distribution in frequency domain}
    \label{fig:power_dist_freq_domain} 
\end{figure*}

Figure \ref{fig:power_dist_freq_domain} illustrates the spectrum of the adversarial noise compared to the average frequency response of power traces without the HTO patch. We notice the adversarial noise contains frequency components that are beyond the expected range of an original raw signal which makes it vulnerable to filtering-based defenses. 

Table \ref{table:filtered_patch_model_accuracy} presents the attack success rate of HTO in the presence of the filter. We couldn't find any frequency range for the filter to retrieve the model accuracy of $AES-T400$ and $AES-T1800$ datasets to the acceptable value. The accuracy of the model for $AES-T500$ and $AES-T600$ datasets partially recovered after applying the filter. We notice that for these two datasets, the model accuracy for the baseline trace and the attacked one is the same while measuring the results by the filter. It means that although the filter impacts the performance of the model, it totally reduces the patching efficiency. On the other hand, the filter significantly degrades the performance of attack for the $AES-T700$ and $AES-T800$ and retrieves the model accuracy to $100\%$.
Moreover, We analyzed the required noise budget to bypass the filter. The results indicate that for all datasets the attacker needs to generate the adversarial patch with a higher amount of power which decreases the stealthiness of the patch.




\begin{table*}[!ht]
\small
    \centering
    \resizebox{\textwidth}{!}{
    \begin{tabular}{| C{2cm} | C{2cm} | C{2cm} | C{2cm} | C{2cm} | C{2cm}  | C{2cm}| C{2cm} |}
             \hline
             \multirow{3}{2cm}{\centering Dataset} & \multirow{3}{2cm}{\centering class} & \multirow{3}{2cm}{\centering Model's accuracy} 
             & \multirow{3}{2cm}{\centering Model's accuracy after patching} & \multirow{3}{2cm}{\centering Model's accuracy after filtering patch}
             & \multirow{3}{2cm}{\centering $[f_{min},f_{max}]$ \\ (MHz)} & \multirow{3}{2cm}{\centering Noise budget (mW)} & \multirow{3}{2cm}{\centering Noise budget to bypass filter (mW)} \\
             &&&&&&&\\
             &&&&&&&\\\hline\hline

             \multirow{2}{2cm}{\centering AES-T400} & Class-0 & 88.9 & 0 & - & \multirow{2}{2cm}{\centering -} & \multirow{2}{2cm}{\centering 0-0.6} & \multirow{2}{2cm}{\centering -}\\\cline{2-5}
                                                    & Class-1 & 97.1 & 0 & - &&&\\\hline

            \multirow{2}{2cm}{\centering AES-T500} & Class-0 & 96.4 & 97.7 & 100 & \multirow{2}{2cm}{\centering [0-7.5]} & \multirow{2}{2cm}{\centering 0-1.2} & \multirow{2}{2cm}{\centering 0-35}\\\cline{2-5}
                                                    & Class-1 & 91.91 & 0 & 47.7 &&&\\\hline

            \multirow{2}{2cm}{\centering AES-T600} & Class-0 & 92.7 & 98.5 & 81.8 & \multirow{2}{2cm}{\centering [0-12.5]} & \multirow{2}{2cm}{\centering 0-1} & \multirow{2}{2cm}{\centering 0-35}\\\cline{2-5}
                                                    & Class-1 & 93.8 & 0 & 32.6 &&&\\\hline

            \multirow{2}{2cm}{\centering AES-T700} & Class-0 & 100 & 100 & 100 & \multirow{2}{2cm}{\centering [0-2.5]} & \multirow{2}{2cm}{\centering 0-1.2} & \multirow{2}{2cm}{\centering 0-11}\\\cline{2-5}
                                                    & Class-1 & 100 & 0 & 100 &&&\\\hline

            \multirow{2}{2cm}{\centering AES-T800} & Class-0 & 100 & 100 & 100 & \multirow{2}{2cm}{\centering [0-5.5]} & \multirow{2}{2cm}{\centering 0-2} & \multirow{2}{2cm}{\centering 0-12}\\\cline{2-5}
                                                    & Class-1 & 100 & 0 & 100 &&&\\\hline

            \multirow{2}{2cm}{\centering AES-T1800} & Class-0 & 99.8 & 0 & - & \multirow{2}{2cm}{\centering -} & \multirow{2}{2cm}{\centering 0-0.5} & \multirow{2}{2cm}{\centering -}\\\cline{2-5}
                                                    & Class-1 & 94.12 & 0 & - &&&\\\hline          
    \end{tabular}
    }
    
    \caption{Model's accuracy for original patches and filtered ones (class-0: benign samples, class-1: HT-inserted samples)}
    \label{table:filtered_patch_model_accuracy}
    
\end{table*}

\noindent
 \textit{\textbf{Adversarial Training:}} Adversarial training (AT) \cite{madry2019deep} is a state-of-the-art defense strategy against adversarial attacks. It can be formulated as follows:
\begin{equation}
\small
    \min _{\theta} \mathbb{E}_{(x, y) \sim \mathcal{D}}\left[\max _{\delta \in B(x, \varepsilon)} \mathcal{L}_{c e}(\theta, x+\delta, y)\right]
\end{equation}

Where $\theta$ indicates the parameters of the classifier, $\mathcal{L}_{c e}$ is the cross-entropy loss, $(x, y) \sim \mathcal{D}$ represents the training data sampled from a
distribution $\mathcal{D}$ and $ B(x, \varepsilon)$ is the allowed perturbation
set. The interpretation of this is that the inner maximization problem is finding the
worst-case samples for the given model, and the outer minimization problem is to train a model robust to adversarial
examples \cite{madry2019deep}.

We used the PGD algorithm \cite{pgd} to solve the inner maximization problem. We set the noise magnitude (epsilon) equal to $0$ with a step size of $0.1$ with a number of iterations equal to $20$. We train the model for around $20$ epochs to reach $99\%$ classification accuracy on clean input samples.

\begin{figure*}[!ht]
\centering
        \begin{subfigure}[b]{0.47\textwidth}
        \centering
            \includegraphics[width=\textwidth]{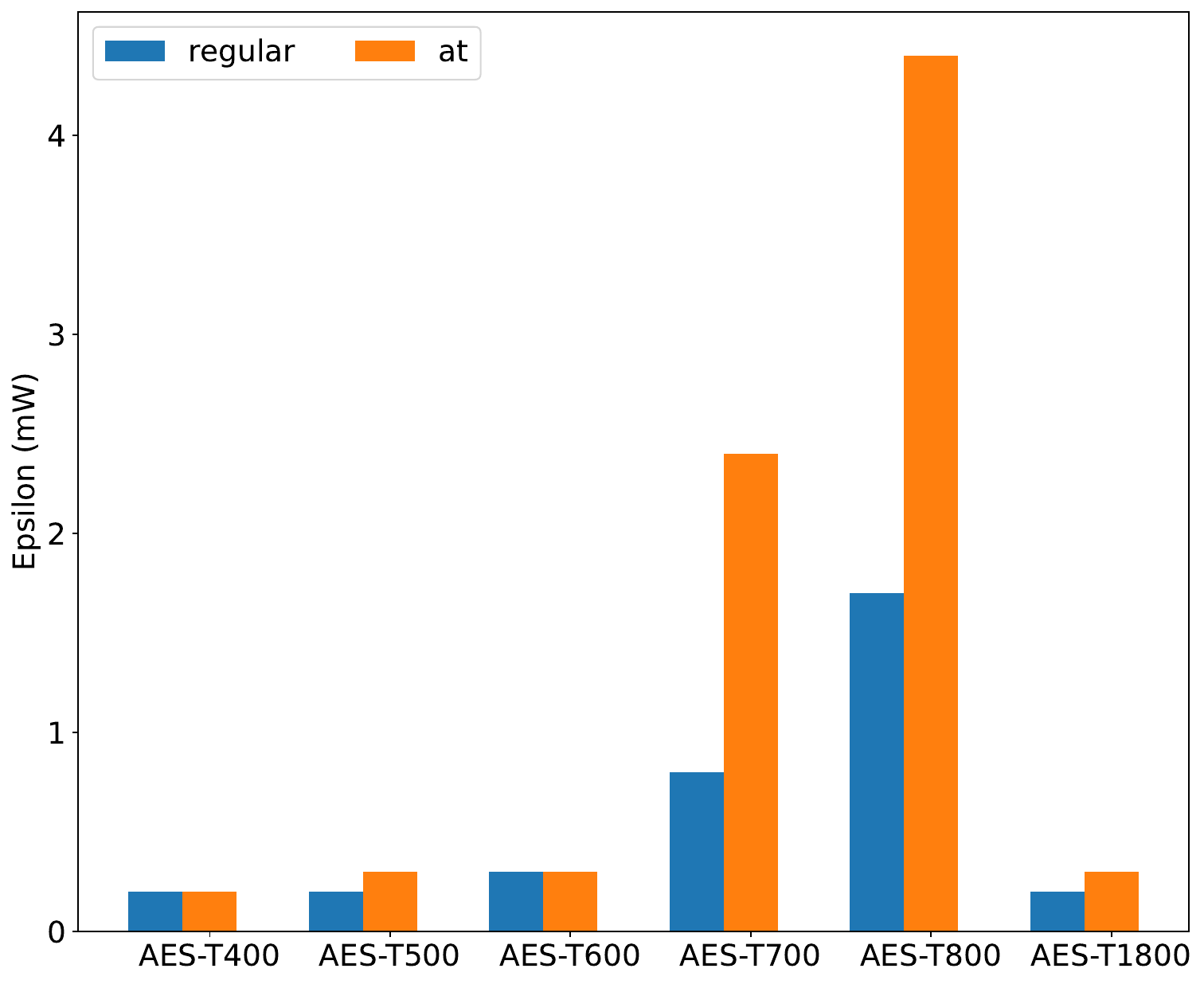}
	       \caption{Synch}
       \end{subfigure}
     \hfill
        \begin{subfigure}[b]{0.47\textwidth}
            \centering
            \includegraphics[width=\textwidth]{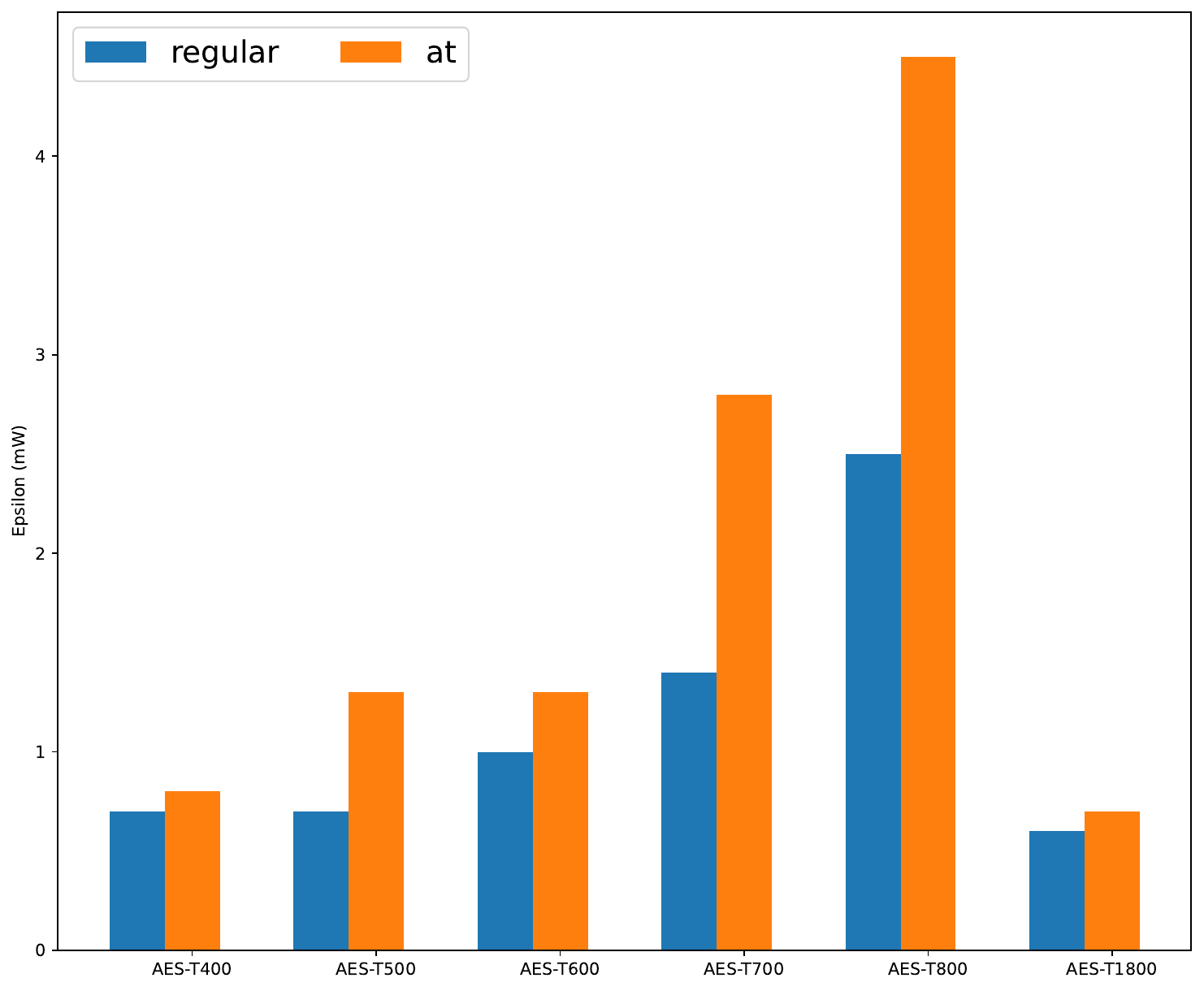}
	       \caption{Unsync}
       \end{subfigure}
       \caption{Impact of patch on baseline and adversarial trained model when the class 1 accuracy drops to $50\%$}
	 \label{fig:AT_eps} 
\end{figure*}

In Figure \ref{fig:AT_eps}, we present a comparative analysis of the effects of patches created using our adversarial patch technique on both AT models and undefended (regular) models . The findings show that AT reduces the efficacy of attacks at lower levels of adversarial noise but becomes less effective against higher noise magnitudes (larger epsilons). This indicates that while adversarial training enhances model robustness against certain attack levels, it is less reliable against more intense adversarial inputs. Additionally, it's important to note that implementing AT, particularly at higher noise levels, as expected, incurs a decrease in baseline accuracy, resulting in a compromise on the overall utility of the model.

\section{Adaptive Attack}\label{sec:adaptive}


The previously adversarial patch generation method considers only a magnitude budget in the time domain without any spectral constraints, which led to easily filtered adversarial impact through a comprehensive filter. 
The objective of an adaptive attacker is to generate adversarial noise in a specific frequency range which corresponds to the expected spectral domain of the designed filter. Consequently, the defender will not be able to cut-off the impact of the injected noise unless with a considerable utility loss. For this reason, we introduce a \textbf{spectral budget} in addition to the noise magnitude budget. Specifically, we not only clip the noise magnitude in the time domain to fit a noise budget, but also project the noise in a subset of the frequency domain to restrain its spectral components. 
The problem can therefore be formulated as follows:
\begin{equation}
\small
\begin{split}
        C(x + shift(\delta, k)) \neq C(x), ~ \forall ~ x \sim \mu_{HT}, s.t. \\
        \left\Vert \delta \right\Vert = \mathcal{D}(x, x+\delta) \leq \varepsilon \\
         ~FFT(\delta) \in [f_{min}, f_{max}]\\
\end{split}
\end{equation}

Where $\varepsilon$ controls the magnitude of the noise vector $\delta$ and 
$shift(.)$ is a function expressed by Equation \ref{eq:shift} that quantifies the adversarial patch signal $\delta$, relatively with regard to a target signal $x$ given an incidence time $k \in [0, d] $ and $k=0$ corresponds to the synchronised case. 
The new constraint on $\delta$ limits the noise frequency components to an acceptable range defined by $f_{min}$ and $f_{max}$, which correspond to the useful frequency range from the defender perspective. Notice that this information is available even in the initial HT threat model.

Algorithm \ref{Algo:adapt} details the noise generation mechanism of our adaptive attack; We include the frequency clipping process within the adversarial noise generation procedure.  In each iteration, after updating the noise, we project the noise back to the target spectral domain using the pass-band filter defined previously. Therefore, at each iteration, we only retain noise samples with the same frequency range as the raw signal. These noise components are then aggregated with the current patch instance.
The shift function is the same as in Algorithm \ref{Algo:unsynch}.

\begin{algorithm}[!ht]
\small
\caption{Adaptive Adversarial Power Trace Generation}
\label{Algo:adapt}
\begin{algorithmic}[1]
\State \textbf{Input:} Data set: $\mathcal{D}$, classifier: $C$,  noise magnitude: $\varepsilon$, standard deviation: $\sigma$, number of iterations: $N$.

\State \textbf{Output:} $\delta$ Adversarial power trace.\\

\State Initialize $\delta \leftarrow rand()$
\While{$iter < N$ }
    \For{each Batch $X_i \sim \mathcal{D}$}\\
            
            \State $\delta \leftarrow \frac{1}{|X_i|} \sum_{j=1}^{|X_i|}
             \{ \{\alpha sign(\nabla_{x} J_{\theta}(C({x_{j} + sh(\delta) }),\ell))\}\}$ \\
            
            \State $\delta \leftarrow Clip_\varepsilon\{ \delta + z \sim N\left(0, \sigma^{2}\right)\} $
 
            \State $\delta \leftarrow Clip_{[f_{min}, f_{max}]} FFT(\delta + z \sim N\left(0, \sigma^{2}\right))$

    \EndFor
\State $iter += 1$
\EndWhile\\



\end{algorithmic}
\end{algorithm}



\section{Related Work}
Several works have been studied in literature to leverage machine learning to detect HTs over the past decade \cite{hasegawa2016hardware, faezi2021htnet}. A novel static method for HT
classification at the design phase based on SVM-based classifier and gate-level
netlist features presented by Hasegawa \etal in \cite{hasegawa2016hardware}. This work was also extended to a similar framework with different ML models, including random
forest (RF) \cite{hasegawa2017trojan} and neural networks \cite{hasegawa2017hardware}. A dataset of side channel signals to train a neural network for Trojan detection at run-time gathered by Faezi \etal \cite{faezi2021htnet}. Noor \etal \cite{noor2017hardware} propose a framework for HT identification based on ML classification with features describing the dominant attributes of HTs. Features such as switching activity and net structure from the gate-level net-list have been extracted, quantified and analyzed to detect malicious activity by Zhou \etal \cite{zhou2016novel}. Xiang \etal \cite{xiang2019random} constructed a random forest classifier for hardware-Trojan detection by extracting the structural and signal characteristics at RTL. An experiment was done to determine the required power and delay SNR for hardware-Trojan identification by lamech \etal \cite{lamech2011experimental}. Inoue \etal \cite{inoue2017designing} present a novel static method for HT classification using gate-level netlist features to identify a part of designed hardware-Trojans. Iwase \etal \cite{iwase2015detection} investigated a new detection technique for HTs based on frequency domain features by converting power consumption waveform data from time into frequency domain. Estimating statistical correlation between the signals in a design and exploring how this estimation can be exploited in a clustering algorithm to identify the HT logic introduced by Cakir \etal \cite{cakir2015hardware}.

Furthermore, ML network can be fooled drastically by an adversarial perturbation  which is not perceptible by human \cite{szegedy2013intriguing}. Generally, the goal of the adversarial attack is to force the well-trained network to predict wrongly with generating the adversarial examples. Adversarial attacks can be categorized into white-box attacks \cite{madry2017towards, carlini2017towards} where the attacker has access to structures and parameters of victim models and black-box attacks \cite{dong2019evading, xie2019improving} where the attacker does not have any information about internal structure of target model and can get a corresponding output only by queries. Although majority of existing attack methods are focusing on multi classification task, it has been exploiting in many fields such as object detection \cite{xie2017adversarial}, face recognition \cite{yin2021adv}, visual tracking \cite{jia2020robust} and etc. Therefore, in this paper, we investigate the ML classifier vulnerabilities introduced by faezi \etal \cite{faezi2021htnet} to design a hardware generating adversarial noise to misclassify the HT-inserted circuits.

\section{Discussion and Concluding Remarks}
In this paper, we shed new lights on the limits of using ML techniques for hardware security. Specifically, we propose a methodology to generate malicious circuity that emulates the consumption of an adversarial power trace to fool the ML-based HT detection classifier. We develop HTO to design circuits that are meant to consume a pre-generated adversarial patch (power trace), for both ASIC and FPGA platforms.

Our results indicate that to reach $100\%$ efficiency, only one transistor circuit might be required for evading ML-based HT detection in the case of ASIC platforms. We also propose two different circuit designs, i.e., ring oscillator-based and DSP-based designs have been proposed to simulate a sneaky adversarial power noise on  FPGA platforms. Our experimental results on a Spartan 6 FPGA board illustrate achieving a $100\%$ efficiency to evade detection is accessible with a maximum of $4$ ROs and $2$ DSPs. Moreover, we introduce spectral-based and adversarial training countermeasures to mitigate the attack's efficiency while keeping the model's performance. We also investigate the impact of adaptive attacks on required power budgets by simultaneously considering the magnitude of the generated patch in the time and frequency domains. 

We believe our work shows a new practical way to exploit ML vulnerability to adversarial noise in a realistic scenario, which is critical for hardware security. While remaining in the classical HT threat model, our approach points out new potential threats that need to be taken into account when using ML techniques as defenses. 

\bibliographystyle{plain}
\bibliography{refs,asplos}

\end{document}